\documentclass[twocolumn,nofootinbib]{revtex4-1}

\usepackage{mathrsfs}
\usepackage{amssymb}
\usepackage{amsmath}
\usepackage{bm}
\usepackage{graphicx}
\usepackage[usenames,dvipsnames]{color}
\usepackage[colorlinks=true,citecolor=blue,linkcolor=magenta]{hyperref}
\usepackage{placeins}
\usepackage{float}

\newcommand{\figpath}{./figures}

\newcommand{\ket}[1]{\vert{ #1 }\rangle}
\newcommand{\bra}[1]{\langle{ #1 }\vert}

\newcommand{\braket}[2]{\langle #1 \vert #2 \rangle}

\begin{document}

\title{Measurement-driven quantum computing: Performance of a {\small 3-SAT} solver}

\author{Simon C. Benjamin}
\affiliation{Department of Materials, University of Oxford, Parks Road, Oxford OX1 3PH, United Kingdom}
\author{Liming Zhao, Joseph F. Fitzsimons} 
\affiliation{Singapore University of Technology and Design, 8 Somapah Road, Singapore 487372}
\affiliation{Centre for Quantum Technologies, National University of Singapore, 3 Science Drive 2, Singapore 117543}

\begin{abstract}
We investigate the performance of a quantum algorithm for solving classical {\small 3-SAT} problems. A cycle of post-selected measurements drives the computer's register monotonically toward a steady state which is correlated to the classical solution(s). An internal parameter $\theta$ determines both the degree of correlation and the success probability, thus controlling the algorithm's runtime. Optionally this parameter can be gradually evolved during the algorithm's execution to create a Zeno-like effect; this can be viewed as an adiabatic evolution of a  Hamiltonian which remains frustration-free at all points, and we lower-bound the corresponding gap. In exact numerical simulations of small systems up to 34 qubits our approach competes favourably with a high-performing classical {\small 3-SAT} solver, which itself outperforms a brute-force application of Grover's search.
\end{abstract}

\maketitle

\section{Introduction}
Quantum computers are capable of performing tasks in a fundamentally different way to any classical computer. It is believed that quantum approaches will yield far more rapid execution times for certain tasks which are important in fields as diverse as chemistry and materials science, security and code-breaking, and machine learning and optimisation. For some of these applications, a quantum computer might achieve an {\it exponential} speedup: if  three problems of increasing complexity need classical machine time of (say) 1 day, 100 days, and 10,000 days respectively, the time needed by a quantum device would only scale polynomially, perhaps even linearly,  as (say) 1, 2, and 3 days. Such problems are obviously very attractive targets for quantum computing, however it can be interesting to explore more modest {\it polynomial} speedups such as the $N\rightarrow\sqrt N$ advantage promised by Grover's algorithm, which in the above example could correspond to 1, 10 and 100 days of runtime. 

In computer science a very famous class of tasks are those called boolean satisfiability ({\small SAT}) problems~\cite{karp1972reducibility}. We will introduce the most famous, {\small 3-SAT} , presently -- but in essence these tasks require one to find a solution that simultaneously satisfies a large number of interrelated constraints. A great many real world problems from logistics to biology can ultimately be related to {\small 3-SAT}. It is not known definitively whether {\small 3-SAT}  can be sped up exponentially with a quantum computer (many researchers suspect not) but even a sub-exponential speedup could be scientifically and commercially important. 

There are several different paradigms for quantum computing. One can be called circuit-based universal computing. Here the qubits are isolated from one another except when the algorithm calls for a precise interaction between them called a `gate operation', the control-{\small NOT} gate is a well-known example. This approach may be seen as the most directly analogous to conventional digital computers, since the algorithm is compiled into a universal set  of low-level operations analogous to conventional logical gates such as {\small AND} and {\small OR}. 

A distinct approach to harnessing quantum effects is adiabatic quantum computing~\cite{Farhi2000}, where qubits interact continuously with their neighbours and thus give rise to a protective energy gap between the correct state of the machine and states where one or more qubits suffer errors. In this approach global parameters such as a magnetic field strength are slowly changed, in order to morph an initial ground state into a computationally interesting final state. This paradigm is technically universal, in the sense that one can in principle translate an algorithm intended for a circuit-based machine into a form suitable for an adiabatic machine~\cite{PhysRevLett.114.140501} -- but in practice it is likely that the adiabatic approach will find its application in specific areas such as quantum annealing. 

Typically, algorithms for circuit-based machines involve applying a series of unitary (reversible) operations on the qubits. Measurements, which are of course irreversible, may be used frequently in the lower-level operation of the machine in order to fight errors or to create useful entanglement. Nevertheless when viewed at a sufficiently high level the algorithm, or at least large portions of it, can be viewed as a unitary evolution. Here we use the term `measurement-driven' to refer to the more unusual cases where the high-level algorithm is a series of measurement events. Note that this is very distinct from `measurement-based' computing~\cite{MBQCbriegel} since that term is used to describe the idea of creating and consuming entanglement via measurements, but in such a way that ultimately the qubits of the algorithm do evolve in a unitary way (up to basis changes).

An early investigation of such a `measurement-driven' approach was published in 2002 by Childs {\it et al}~\cite{ChildsPRA2002} and considers how a circuit-based machine might mimic the evolution of an adiabatic computer. The approach treats the Hamiltonian of the adiabatic machine as a measurement to be made on a circuit-based computer. This measurement is made repeatedly and each time the Hamiltonian is changed slightly. The Zeno effect dominates the resulting dynamics provided that the changes are sufficiently small -- then, assuming the initial measurement yielded a result corresponding to the ground state energy, there is a high probability that each of the following measurements again yields the (instantaneous) ground state. In this way the system is driven through a series of small steps to reach the ground state of a final Hamiltonian that is very different to the first. The authors relate the success probability of the measurement-drive process to the size of the energy gap between the ground and excited states in the Hamiltonian; this same gap would determine the runtime of a real adiabatic machine, and thus the authors were able to make a concrete comparison. 

An impediment to realising the approach of Ref.~\cite{ChildsPRA2002} is the complexity of measuring the entire Hamiltonian of a system. In a recent paper we revisit this scenario, providing a new proof that one can perform the same measurement-driven evolution simply by measuring the individual terms of the Hamiltonian~\cite{Liming2017}. The approach of repeatedly measuring terms of a Hamiltonian works when the Hamiltonian is frustration free, and is related to the previously-studied topic dissipation-driven computing and (especially) state preparation~\cite{VerstraeteDissip}.

The present document describes and assesses a measurement-driven algorithm that seeks the solutions to {\small 3-SAT} problems. The algorithm was previously described by one of us (SCB) in a notes posted online~\cite{Simon3SAT2015} in 2015; here we provide a fuller analysis, with proofs of monotonic convergence to the classically-meaningful solution(s), a lower bound on the related Hamiltonian gap, and more detailed numerical studies of performance. The algorithm is conceptually very simple: we perform a repeating cycle of measurements which correspond directly to the structure of the {\small 3-SAT} problem. The measurement outcomes are labelled `{\small TRUE}' and `{\small FALSE}' and if the latter ever occurs then the algorithm restarts. We show that if the {\small 3-SAT} problem has a single satisfying solution then our algorithm proceeds monotonically toward a final state that encodes the solution; given multiple classical solutions, we tend to a corresponding subspace. In numerical simulations we find that this very na\"ive approach achieves a similar (and somewhat superior) performance to the most well-studied high performing classical algorithm, and appears to have superior scaling. The classical algorithm itself has provably superior performance to a direct use of Grover's algorithm to search for the {\small 3-SAT}'s solution(s).

It is worth reiterating that the algorithm discussed here aims to solve the classical {\small 3-SAT} problem by quantum means, it does not aim to solve quantum {\small 3-SAT} which is a generalisation of the {\small 3-SAT} problem itself. The latter, quantum {\small k-SAT}  was introduced by Bravyi~\cite{Bravyi06} with a focus on $k=2$ terms, and remains an active area of research (see e.g.~\cite{Farhi2016,Sattath2016}). We presently remark on the possible overlaps.

\section{Satisfiability}
Boolean satisfiability problems involve a `proposition' $P(b_1,b_2,...b_n)=\ ${\small TRUE} where the $n$ boolean variables $b_i$ may each take the value {\small TRUE} or {\small FALSE}. Typically the challenge is to determine whether any choice of values for the variables $b_i$ can satisfy the formula.

In the particular case called 3-satisfiability ({\small 3-SAT}) the expression $P(b_1,b_b,...b_n)$ is formed by {\small AND}'ing together a number of {\it clauses}, and each clause involves three variables {\small OR}'ed together. 

\smallskip
\noindent{\bf Example:}\ $(b_1 \lor \lnot b_2 \lor b_3) \land (\lnot b_1 \lor b_4 \lor b_5) \land ... = {\rm TRUE}.$\\

Here the symbol  $\land$ denotes the logical {\small AND}, $\lor$ denotes the logical {\small OR}, while  $\lnot$ denotes the logical {\small NOT} which is also referred to as {\em negation}. Table~\ref{littleLogicTable} summarises these standard elements. Note that {\it verifying} a candidate solution merely involves  checking that each independent clause is {\small TRUE}; if-and-only-if they all are, the {\small 3-SAT} is indeed satisfied by that candidate solution. In this paper we will use the symbol $m$ to refer to the number of clauses in a \small{3-SAT}, and $n$ to refer to the number of booleans.

\begin{table}[t]
\centering
\begin{tabular}{|cc|c|c|c|}
\hline
$\ \ x\ \ $ & $\ \ y\ \ $  & $\ \ x \land y\ \ $ & $\ \ x \lor y\ \ $ & $\ \ \lnot x\ \ $ \\
\hline
F & F & F & F & T\\
F & T & F & T & T\\
T & F & F & T & F\\
T & T & T & T & F\\
\hline
\end{tabular}
\caption{Table defining operations {\small AND}, {\small OR} and {\small NOT}.}
\label{littleLogicTable}
\end{table}

\section{Classical {\small 3-SAT} Solvers}
The task of efficiently solving  {\small 3-SAT} instances is a long studied challenge in computer science. It is believed that any algorithm will require a running time which is exponential in $n$~\cite{wikiETH}. Clearly the problem can be solved by a brute force search over all $2^n$ variable assignments. Then each failed test of a potential assignment prior to the solution will require some portion of the $m$ clauses to be checked, so that $t=m\,2^n$ is a trivial upper bound to the number of clause checks required. Throughout this paper we take the {\bf expected number of clause checks} to be the metric for the running time of an algorithm.

The na\"ive classical bound can be dramatically improved upon. A series of publications have succeeded in reducing the base $K$ in the the expression for the expected running time, $t\propto K^n$. The current record holder appears to be an algorithm due to Paturi {\it et al}~\cite{PaturiImproved2005} which boasts an upper bound of $K=2^{2\ln 2-1}\approx1.307$. A earlier paper by Sch\"oning~\cite{SchoningOrig} has a marginally worse upper bound of $K=1.334$ but is considerably more simple making direct comparison with the quantum case more straightforward. Therefore we base our comparison with classical performance on Sch\"oning's algorithm which we specify in Box 1.

Notice that the algorithm chooses a random point in the space of boolean assignments at step (\ref{newTry}) and then takes a partially-random walk by flipping one boolean at time, such that a given {\small FALSE} clause will become {\small TRUE} (but other clauses that were {\small TRUE} may become {\small FALSE}). Note that we should randomly find a failed clause among all the failed clauses, since this randomness considerably improves performance over maintaining a fixed check-order.

If a full solution is not found after $c_\text{max}$ such changes, the algorithm restarts. Sch\"oning motivates this by noting that it limits the problem of the candidate state tending to move away from the solution; by introducing the constraint he bounds the probability that a given random walk indeed finds the solution. Using $c_\text{max}=3n$ his analysis concludes that the runtime is within a polynomial factor of $(4/3)^n=1.334^n$ for {\small 3-SAT} (actually $\{2(1-1/n)\}^n$ for {\small $k${\small -SAT}}).
However we find that, for the \small{3-SAT} problems we consider here (ranging from 20 to 34 booleans) the statistical performance of the classical algorithm improves, both in mean runtime and in variance, when $c_\text{max}\rightarrow\infty$. Thus we make this adjustment.

\vspace{4pt}
\setlength{\fboxsep}{8pt}
 \noindent\fbox{\parbox[c][6.0cm][c]{.9\linewidth}{
\smallskip
{\bf Box 1: Sch\"oning's algorithm}
\begin{enumerate}
  \item \label{newTry}Select a random assignment for all $n$ booleans, and 
   set a counter $c=1$.
  \item   \label{checkAll} Systematically check all $m$ clauses {\it in random order}, checking the truth value of each until either
\begin{enumerate}
\item All clauses have been checked and are {\small TRUE}. A satisfying solution has been found, so exit.
\item Some clause is found to be  {\small FALSE}.
\end{enumerate}
  \item If $c>c_\text{max}$ then goto (\ref{newTry}). 
  \item For the clause found to be {\small FALSE}, select one of its three booleans at random and invert it.
  \item Increment counter $c$ and goto (\ref{checkAll}).
\end{enumerate}
}}

\section{A Quantum {\small 3-SAT} Solver}

A simple application of Grover's search algorithm would yield a quadratic improvement over the brute force classical search, from $t\propto 2^n$ to $t\propto (\sqrt 2)^n$ i.e. $K=1.41$, but this is in fact above the upper bound on the best classical solutions as noted above. It has been suggested~\cite{Ambainis} that one of the high performing classical algorithms can be accelerated by replacing the random search with a Grover-type coherent evolution, leading to a hybrid algorithm with expected running time bounded by $\propto (1.153)^n$.

Here we describe a different quantum approach based on repeatedly making projective measurements, each corresponding to evaluating the truth value of a generalised clause. One might naturally associate $\ket{1}$ with {\small TRUE} and $\ket{0}$ with {\small FALSE}, and we generalise this to consider non-orthogonal qubit states associated with the parameter $\theta$ as follows
\begin{eqnarray}
\text{\small TRUE}\rightarrow\ket{\theta}&=&R_Y(+\theta)\,\ket{+}=\cos\alpha\ket{0}+\sin\alpha\ket{1}, \nonumber\\
\text{\small FALSE}\rightarrow\ket{\bar\theta}&=&R_Y(-\theta)\,\ket{+}=\sin\alpha\ket{0}+\cos\alpha\ket{1}. 
\label{eqn:TRUEandFALSE} 
\end{eqnarray}
with $\alpha=(\tfrac{\pi}{4}+\tfrac{\theta}{2})$, and the rotation operator
\begin{equation}
R_Y(\theta)=\left(\begin{array}{cc}\cos\frac{\theta}{2} & -\sin\frac{\theta}{2}  \\ \sin\frac{\theta}{2} & \cos\frac{\theta}{2} \end{array}\right).
\label{eqn:defY}
\end{equation}
When $\theta=\tfrac{\pi}{2}$, the {\small TRUE} and {\small FALSE} states correspond to the classical limits, i.e. $\ket{\theta}=\ket{1}$ and  $\ket{\bar\theta}=\ket{0}$. However when $\theta=0$, the states are degenerate: $\ket{\theta}=\ket{\bar\theta}=\ket{+}$. 

\begin{figure}[b!]
\centering
\includegraphics[width=.99\linewidth]{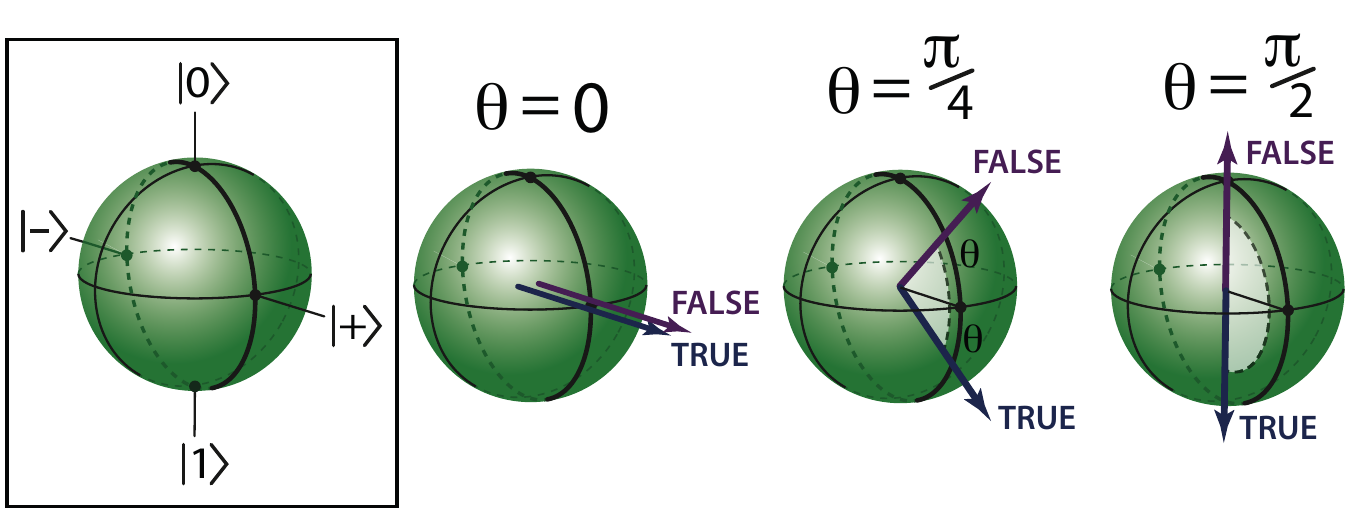}
\caption{Examples of `{\small TRUE}' and `{\small FALSE}' according to $\theta$.  }
\label{fig:alg_sculpt}
\end{figure}

\noindent We also introduce
\begin{equation}
\ket{\theta_\perp}=R_Y\left(\pi+\theta\right)\ket{+}\ \ \ \ {\rm and}\ \ \ \ \ket{{\bar \theta_\perp}}=R_Y\left(\pi-\theta\right)\ket{+}.
\label{eqn:orthogStates}
\end{equation}
These are the states which are orthogonal to $\ket{\theta}$ and $\ket{{\bar \theta}}$ respectively, i.e. 
\begin{equation}
\braket{\theta_\perp}{\theta}=0\ \ \ \ \ {\rm and}\ \ \ \braket{{\bar \theta}_\perp}{{\bar \theta}}=0.
\end{equation}
However note that the states $\ket{\theta}$ and $\ket{\bar\theta}$ are not orthogonal to one another given any $\theta$ that lies {\it within} the range $0<\theta<\tfrac{\pi}{2}$ (note the inequalities are strict). For such a $\theta$, we find
\begin{equation}
\braket{\theta}{\bar\theta}\neq 0,\ \ \ \braket{\theta_\perp}{\bar\theta_\perp}\neq 0,\ \ \ \braket{\theta_\perp}{\bar\theta}\neq 0,\ \ \ \braket{\bar\theta_\perp}{\theta}\neq 0.
\end{equation}

The quantum algorithm is based around performing a series of measurements. Each measurement corresponds to a clause in the classical {\small 3-SAT}, and acts on three qubits $q_i$, $q_j$, $q_k$ which correspond to the three booleans $b_i$, $b_j$, $b_k$ in the clause. Considering the classical clause, we know that there is one specific assignment of values to the booleans for which the clause will evaluate to {\small FALSE}. The quantum measurement is effectively checking that the qubits are {\it not} in the corresponding quantum state. Consider the product state of three qubits $\ket{ABC}$ where 
\begin{eqnarray}
\ket{A}&=&\ket{\theta_\perp}\ \text{if}\ b_i\rightarrow\text{{\small TRUE} would satisfy the clause,}\nonumber\\
\ket{A}&=&\ket{\bar\theta_\perp}\ \text{if}\ b_i\rightarrow\text{{\small FALSE} would satisfy the clause.}\nonumber
\end{eqnarray}
and define $\ket{B}$ and $\ket{C}$ analogously for booleans $b_j$ and $b_k$ respectively. We wish to measure whether or not the three qubits are in state $\ket{ABC}$. This determination can be made using a triply-controlled {\small NOT} gate acting on an ancilla which is then measured, with suitable single qubit rotations applied before and after. In the local space of the three qubits, our projective measurement will either `collapse' to  $\ket{ABC}$, or to some state in the orthogonal seven-dimensional subspace. In the former case where we say the clause check {\it failed} and we restart the algorithm. In the latter case we say the clause check is `passed', and we know then that the complete state $\ket{s}$ of the quantum register satisfies
\begin{equation}
\mathcal{C}_i\ket{s}=\ket{s}
\label{eqn:clauseCheckStatement}
\end{equation}
where
\begin{equation}
\mathcal{C}_i\equiv\mathcal{I}-\mathcal{P}_i\ \text{and}\ \mathcal{P}_i=\ket{ABC}\bra{ABC}
\label{eqn:CiDef}
\end{equation}
where $\mathcal{I}$ is the identity.

For example, if a clause in the \small{3-SAT} is $(b_i\lor b_j \lor \lnot b_k)$, i.e. ``$b_i$ is {\small TRUE}, or $b_j$ is {\small TRUE}, or $b_k$ is {\small FALSE}'', then the corresponding projector for the state we wish to exclude is 
\[
\mathcal{P}= \ket{\theta^i_\perp\ \theta^j_\perp\ {\bar \theta}^k_\perp}\bra{\theta^i_\perp\ \theta^j_\perp\ {\bar \theta}^k_\perp}.
\]
This is of course the projector for the state formed from the three states that are each orthogonal to ``what the clause wants''. 

The problem thus defined is a restricted form of the general quantum {\small 3-SAT} problem~\cite{Bravyi06} -- in such a problem one might specify any set of projectors, each involving three qubits but not in general in product form. Here, our projectors are indeed in product form and moreover a given qubit is referenced by a projector only by $\ket{\theta_\perp}$ or by $\ket{\bar \theta_\perp}$. 

\vspace{4pt}
\setlength{\fboxsep}{8pt}
 \noindent\fbox{\parbox[c][6.7cm][c]{.9\linewidth}{
\smallskip
{\bf Box 2: The quantum algorithm}
\begin{enumerate}
  \item \label{newTry}Initialise all qubits to $\ket{+}$. Set counter $c=1$.
  \item \label{clauseCycle} Sequentially perform a {\it clause check cycle}, performing a clause check operation $\mathcal C_i$ for each of the $m$ clauses as per Eqn.~(\ref{eqn:clauseCheckStatement}) until either
  \begin{enumerate}
\item The projector for each clause has been applied successfully, or
\item Some clause fails the projector.
\end{enumerate}
  \item If a clause failed, then goto (\ref{newTry}). 
  \item \label{endOfCyc} If $c<c_Q$ then increment $c$ and optionally update $\theta$, now goto (\ref{clauseCycle}).
  \item \label{lastStepQ} No clause has failed and $c=c_Q$ thus we have performed a successful run; measure all qubits. \\ 
  If final $\theta<\tfrac{\pi}{2}$ we may require more information; if so goto (\ref{newTry}).  
\end{enumerate}
}}
\vspace{4pt}

Notice that when parameter $\theta$ is strictly within the limits $0<\theta<\tfrac{\pi}{2}$ then certain clause check operations will not commute. Specifically, when two clauses $A$ and $B$ in the {\small 3-SAT} both refer to a given boolean, but one clause `wants' the boolean to be {\small TRUE} while the other wants {\small FALSE}, then the corresponding clause check operations will not commute. This means that if the quantum state passes clause check $A$, and then subsequently passes clause check $B$, the resulting state would not necessarily pass clause check $A$ if we rechecked it. The act of performing the latter check has disrupted the system's state with respect to the former check. Consequently it is not trivial to describe the net effect of applying all $m$ clause checks in a cycle. (It is interesting to note that the effect of non-commuting operators in the general quantum {\small 2-SAT} problem has recently been explored in the context of proving a quantum generalisation of Lov{\`a}sz Local Lemma~\cite{Sattath2016}.)

Despite the non-commuting operators, if the classical {\small 3-SAT} has one or more satisfying solution(s) then one can write a set of independent state(s) which have the property that they will pass all clause checks with certainty. That is, there exist state(s) $\ket{\bm \theta}$ for which
\begin{equation}
{\mathcal C}_i\ket{\bm \theta}=\ket{\bm\theta}\ \ \forall i\ \ \ \ \Leftrightarrow\ \ \ \ {\mathcal P}_i\ket{\bm\theta}=0\ \ \ \forall\,i.
\label{eqn:stableState}
\end{equation}

\begin{figure}[b!]
\centering
\includegraphics[width=.95\linewidth]{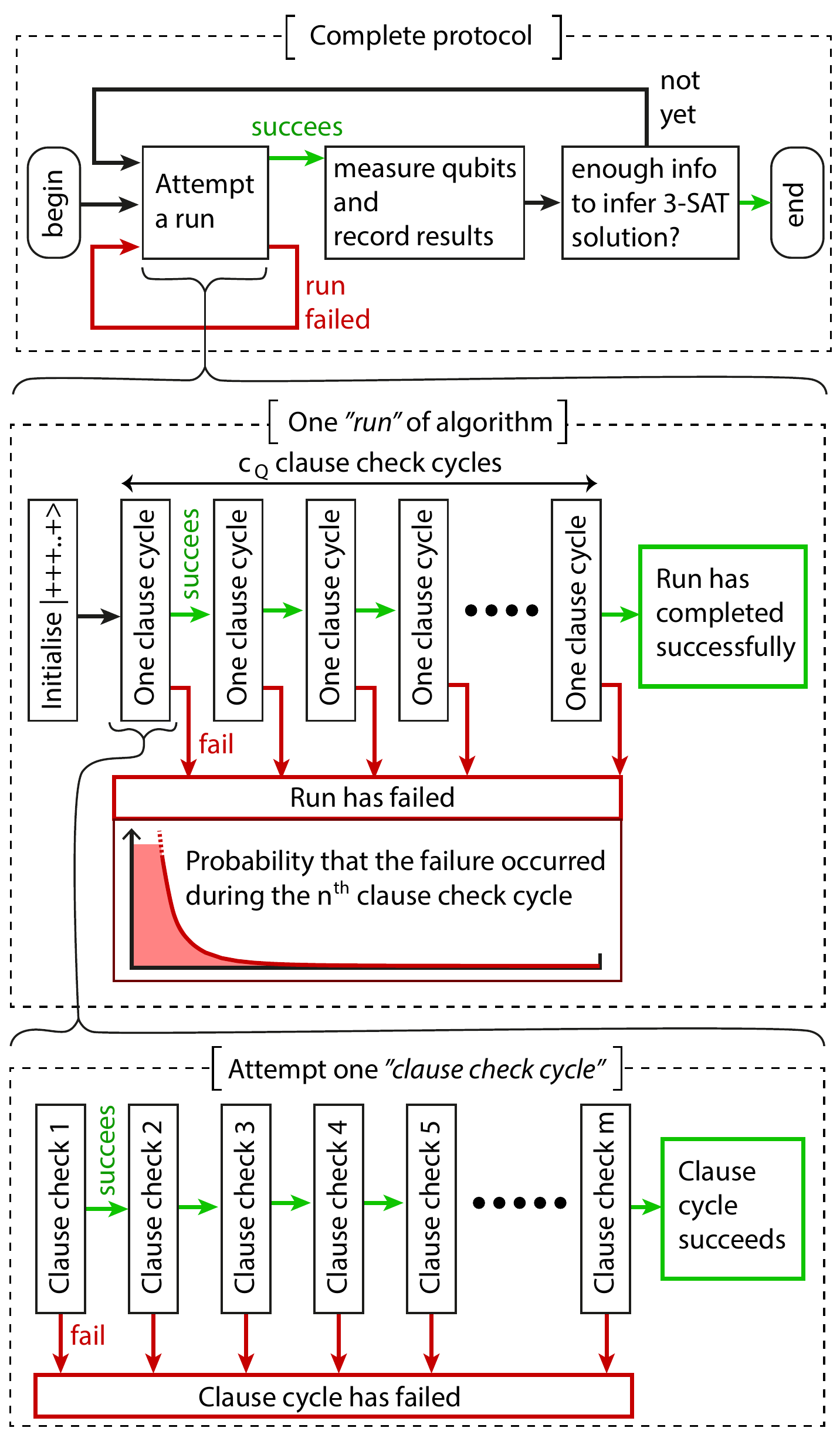}
\caption{ The quantum algorithm.  }
\label{fig:alg_sculpt}
\end{figure}

We can create one such state $\ket{\bm\theta}$ for each solution of the classical {\small 3-SAT}. We prepare a product state where the $i^{th}$ qubit is in state 
\begin{eqnarray}
\ket{q^ i}&=&\ket{\theta}\ \text{if}\ b_i\rightarrow\text{{\small TRUE} in the satisfying solution,}\nonumber\\
\ket{q^i}&=&\ket{\bar\theta}\ \text{if}\ b_i\rightarrow\text{{\small FALSE} in the satisfying solution.}\nonumber
\end{eqnarray}
Equivalently, we can write the entire quantum state as
\begin{equation}
\ket{\bm \theta}\ =\ \bigotimes_i\ket{q^i}\ =\ \bigotimes_i \ R_Y\left(L_i\theta\right)\ket{+}_i 
\label{eqn:solutionStateTheta}
\end{equation}
where $L_i$ is $+1$ if $b_i\rightarrow${\small TRUE} in the satisfying solution, and $L=-1$  if $b_i\rightarrow${\small FALSE}.

One can verify immediately that this state meets the condition in Eqn.~(\ref{eqn:stableState}) since at least one qubit will be orthogonal to one of the three single-qubit projectors forming $\mathcal P_i$, for all $i$. We refer to a stable state $\ket{\bm\theta}$ as a solution state. Our quantum algorithm is designed to filter an initial state until we are left in the solution state, if the classical {\small 3-SAT} has a unique solution, or in a subspace spanned by the set of solution states $\{\ket{\bm \theta}\}$ if there are multiple solutions. 

Notice that although the quantum and classical algorithms have apparent similarities, they are quite different. The classical algorithm repeatedly finds that its candidate solution is incorrect, and attempts to modify that candidate to discover a satisfying solution. Meanwhile the quantum algorithm repeatedly checks that the present state does satisfy clause checks, and restarts if any clause check fails. 

It is important to exclude the possibility there are other states, besides those of the form Eqn.~(\ref{eqn:solutionStateTheta}), which also meet the condition in Eqn.~(\ref{eqn:stableState}) and are therefore stable endpoints to the quantum algorithm. In fact there are no such states: The dimension of the space of stable states $\ket{s}$ is equal to the number of satisfying solutions to the classical {\small 3-SAT}. This is what one expects intuitively but establishing it rigorously takes some space and so we defer it to Appendix~\ref{appendix:NoOtherSolutions}. Consequently the fidelity of the instantaneous state $\ket{s}$ with respect to the solution state(s) is a strictly increasing monotone as we successively perform clause check cycles, i.e. the quantum algorithm cannot be `trapped', see Appendix~\ref{appendix:FidelityMonotonic}.

If the quantum algorithm is run with a fixed $\theta<\tfrac{\pi}{2}$ then it will asymptote to a steady state $\ket{\bm \theta}$ (or to the space of such states when there are multiple solutions) but it will never perfectly reach it, so we must choose to halt and measure at some point. Assuming that we have run the quantum algorithm for long enough that the actual state $\ket{s}$ of the register is a close approximation to solution state $\ket{\bm \theta}$, then when we measure $\ket{s}$ in the $z$-basis each qubits will give us the `correct' classical solution, 0 or 1, with probability close to $\sin^2\alpha=\tfrac{1}{2}+\tfrac{1}{2}\sin\theta$. Thus we gain partial information about the satisfying solution; we may need to run the algorithm a number of times before we can infer the full solution with good probability. It is the problem of inferring whether a classical coin, with a known degree of bias, is baised towards heads or tails; as noted in Appendix~\ref{appendix:InfoFromThetaBelowMax} this is efficient even for small bias. We call this the fixed-theta approach.

Alternatively we may opt to slowly change $\theta$ in such a way that it reaches $\theta=\tfrac{\pi}{2}$ after a defined number of clause checks, at which point we halt and measure for the complete solution. We call this the evolving-theta approach.

For simplicity we will restrict our numerical study to the basic algorithm just described, but there are clearly several possible variants and enhancements to the basic quantum algorithm. For example, one could consider schemes to introduce qubits progressively, enforcing a growing subset of all clauses. One might also consider the possibility of reinitialising qubit(s) associated with a failed clause, e.g. to a randomly chosen state, in the spirit of dissipation-driven approaches~\cite{VerstraeteDissip}. Another possibility is to explore the use of weak measurements to manipulate the success probability of a clause check, along the lines of Ref.~\cite{Sattath2016}. 

\section{Formulation as a Hamiltonian, and gap estimate}

It is interesting to cast the present algorithm into the form of a Hamiltonian problem. We first make a slight adjustment to algorithm as specified in Box 2: Instead of completing `clause check cycles' systematically, we select and perform each successive clause check operation uniformly at random. This `random ordering' variant proves to have similar (albeit somewhat slower) performance to the ordered variant if we assess it in numerical simulation; but it is fundamentally the same approach with the same promise of monotonically increasing fidelity on the solution state $\ket{\bm \theta}$ as specified in Eqn.~(\ref{eqn:solutionStateTheta}). We can then write a Hamiltonian
\begin{equation}
\mathcal{H} = \frac{1}{m} \sum_i \mathcal{P}_i
\label{eqn:Hamiltonian}
\end{equation}
such that state $\ket{\bm \theta}$ is the ground state with zero energy (the inclusion of the prefactor $m^{-1}$ restricts the energy spectrum to the range $0\leq E\leq1$). The Hamiltonian is `frustration free' in the sense that the expected energy of each individual term of $\mathcal{H}$ is  minimised by the ground state $\ket{\bm \theta}$. Because it is frustration free, we could employ dissipation to prepare state $\ket{\bm \theta}$ as in Ref.~\cite{VerstraeteDissip}, or a sequence of measurements of $\mathcal{P}_i$ with a gradually evolving $\theta$ value to synthesise an adiabatic-like controlled evolution as envisaged in Ref.~\cite{Liming2017}. In both cases, a lower bound on the performance can be obtained by lower bounding the gap between the ground state of $\mathcal{H}$ (or its degenerate ground state manifold if the classical {\small 3-SAT} has multiple solutions) and the lowest excited state. We know from Appendix~
\ref{appendix:gapDerivation} that the gap is finite for all $0<\theta\leq \frac{\pi}{2}$. In Appendix~\ref{appendix:NoOtherSolutions} we present the derivation of the following lower bound on the gap:
\begin{align*}
\bra{\Phi}\mathcal{H}\ket{\Phi} &\geq \frac{1}{m}  \sin(\theta)^6 \left(\frac{1-\cos(\theta)}{1+\cos(\theta)}\right)^n,
\end{align*}
where in this expression $\ket{\Phi}$ is understood to be a state orthogonal to the ground state $\ket{\bm \theta}$. 

The bound is found to be loose by comparison to numerically tractable instances with $n$ ranging from $20$ to $34$ as described in the next section; whether it is tight or loose in the large $n$ limit is not known. If indeed the bound becomes tight, then the quantum algorithm may become equivalent to a classical random search since the gap expression is trivially maximised by $\theta=\frac{\pi}{2}$, the classical limit of the allowed range $0\leq\theta\leq\frac{\pi}{2}$. The trend of the performance of the quantum algorithm over the range $n=20$ to $n=34$ suggests otherwise, i.e. that the quantum algorithm becomes increasingly superior to classical approaches as $n$ increases. It is interesting to ask whether a tighter bound could be derived analytically. 

\section{Numerical simulations}

We now discuss the performance of both the classical and quantum algorithms on a range of {\small 3-SAT} problems with up to $34$ booleans. These {\small 3-SAT} problems were generated randomly and postselected on the number of satisfying solutions (usually exactly one) as described in Appendix~\ref{appendix:3SATgeneration}.

\bigskip
\noindent {\bf Performance with fixed $\theta$}

\begin{figure*}
\centering
\includegraphics[width=1\linewidth]{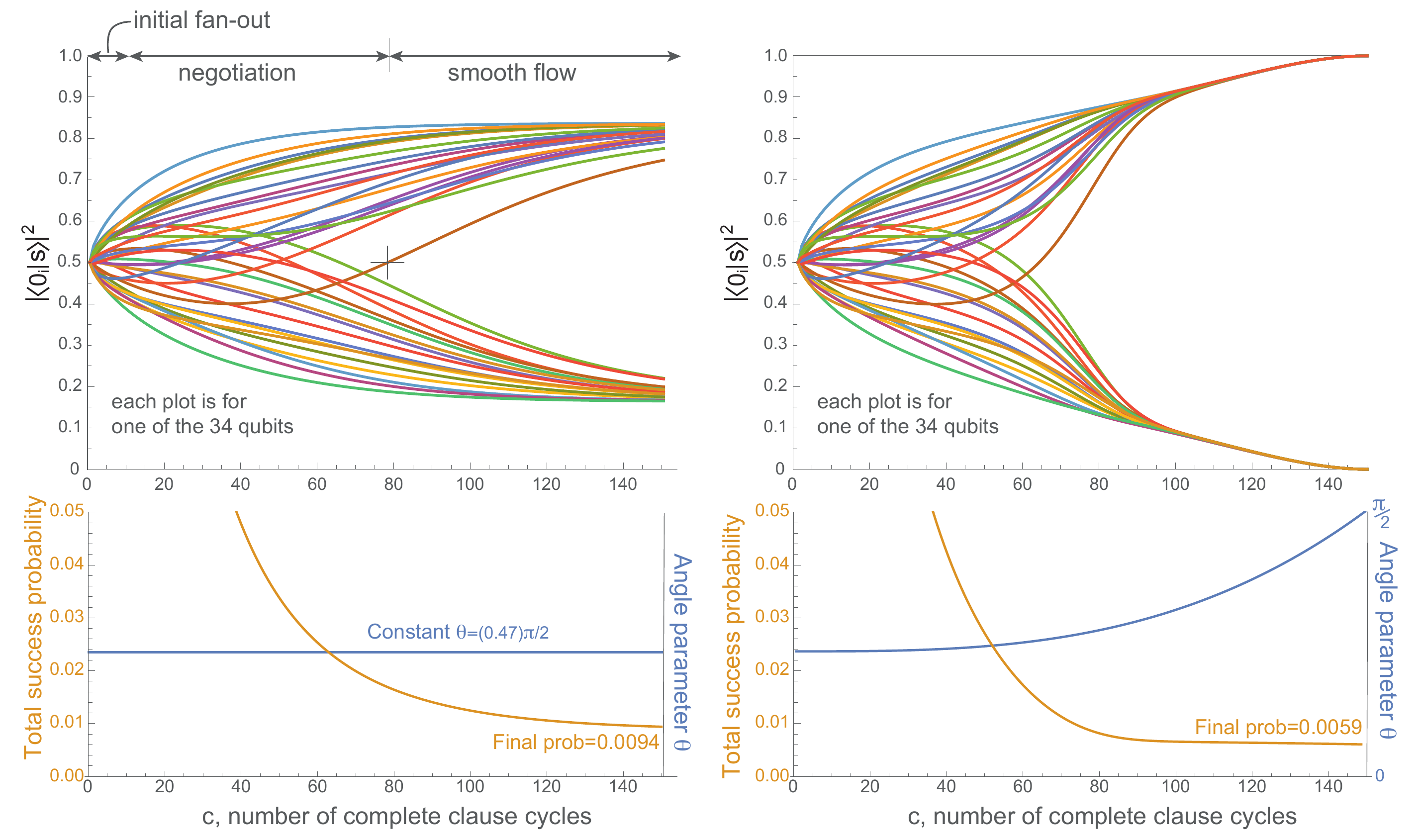}
\caption{Showing how the observables evolve {\it within} one run of the quantum algorithm given a certain $34$-boolean {\small 3-SAT}. On the left, the algorithm is run with fixed $\theta$. On the right, the algorithm is run with an evolving $\theta$. }
\label{fig:34qubits}
\end{figure*}

\smallskip
 It is interesting to plot the progress that each qubit makes toward $\ket{0}$ or $\ket{1}$ as the algorithm progresses. In our simulation software we do this by pausing at the end of each cycle of clause evaluations, at step (\ref{endOfCyc}), and for each qubit in turn we evaluate the probability that it would be found in state $\ket{1}$ if it were measured. However we do not measure it, but rather we continue the algorithm. A plot showing the behaviour for a typical {\small 3-SAT} of 34 booleans is shown in Fig.~\ref{fig:34qubits}, left panel. In this example the classical {\small 3-SAT} has exactly one satisfying solution; this is generally the case for the numerical simulations we show unless stated otherwise.

In Fig.~\ref{fig:34qubits} we have specified a total of $c_Q=150$ clause cycles.  The register starts in state $\ket{+}^{\otimes 34}$ and therefore initially all qubits have equal probability associated with $\ket{0}$ and $\ket{1}$. Over the first few clause cycles, each qubit develops a bias toward or away from $\ket{1}$ which is principally determined by the proportion of clauses which `want' the corresponding boolean to be {\small TRUE}. However after this initial {\it fan out}, we enter a period which is marked as {\it negotiation} during which the bias of certain qubits inverts. Eventually all qubits have the `correct' bias in the sense that each favours either  $\ket{1}$ or $\ket{0}$ according to the corresponding boolean being {\small TRUE} or {\small FALSE} in the unique satisfying solution. 
\begin{eqnarray}
|\braket{1_i}{s}|^2&>&0.51\ \ \text{if boolean}\ b_i=\text{\small TRUE} \nonumber\\
|\braket{0_i}{s}|^2&>&0.51\ \ \text{if boolean}\ b_i=\text{\small FALSE}. \nonumber
\end{eqnarray}
We define the number of clause check cycles needed to reach this point as $c_\text{smooth}$. Eventually, with increasing number of clause check cycles the bias will asymptote to $\sin^2\alpha=\tfrac{1}{2}+\tfrac{1}{2}\sin\theta$ as the state of the system $\ket{s}\rightarrow\ket{\bm\theta}$. If we halt the algorithm and measure the qubits after we have performed more than $c_\text{smooth}$ cycles, we will gain partial information about the classical solution. If we specify a target number of clause cycles $c_Q$ so that indeed $c_Q>c_\text{smooth}$ then repeatedly generating this state and measuring will eventually lead us to infer the classical solution. In practice we will not know $c_\text{smooth}$ but we can estimate it and, if we do not reach a classical solution, we can try again having extended it. A more sophisticated approach would be to engineer  the initial product state so that it already favours the current best hypothesis for the classical solution. However, in numerical simulations it appears that a yet more efficient approach is to perform an adiabatic-like evolution as we now explain.

\bigskip
\noindent {\bf Performance with evolving $\theta$}
\smallskip

 In the right-hand panel of Fig.~\ref{fig:34qubits} we show the results of an approach where the internal parameter $\theta$ is allowed to vary over the course of the algorithm. Initially it is set to the value $\theta_\text{init}=(0.47)\tfrac{\pi}{2}$ which was the constant used for the left-hand panel. Therefore the initial behaviour in the qubit biases is similar. However, we then begin to increase $\theta$ steadily until we reach $\theta=\tfrac{\pi}{2}$. In fact the form is 
 \begin{equation}
 \theta(\text{cycle}\ c)= \theta_\text{init}+(\tfrac{\pi}{2}-\theta_\text{init})\left(\frac{c}{c_Q}\right)^3.
 \label{eqn:thetaEvolve}
 \end{equation}
Consequently, the biases eventually reach $100\%$ and one can read out the classical {\small 3-SAT} solution simply by measuring the register without any need for repetition. Importantly, this has relatively little effect on the success probability -- it falls from $0.0094$ for the fixed-$\theta$ variant to $0.0059$ for the new variant, with the great benefit that we discover the full classical solution in `one shot'.

In the fixed-$\theta$ approach (left side Fig.~\ref{fig:34qubits}) it is clear that $c_\text{smooth}$ the number of cycles needed to reach the point where all biases are `correct' is a useful metric for the running time of the algorithm. In fact the same is true of the evolving-$\theta$ approach, because we find that the total success probability falls significantly if we begin our adiabatic evolution of $\theta$ `too soon', i.e. the system follows the evolving $\theta$ with high probability only if we are past the `negotiation' phase. We can regard $c_\text{smooth}$ as one measure of the algorithm's efficiency.  In Appendix~\ref{appendix:cSmooth} we plot the average $c_\text{smooth}$ for a series of cases where the number of booleans varies from $20$ to $30$. 
 
\begin{figure}
\centering
\includegraphics[width=1\linewidth]{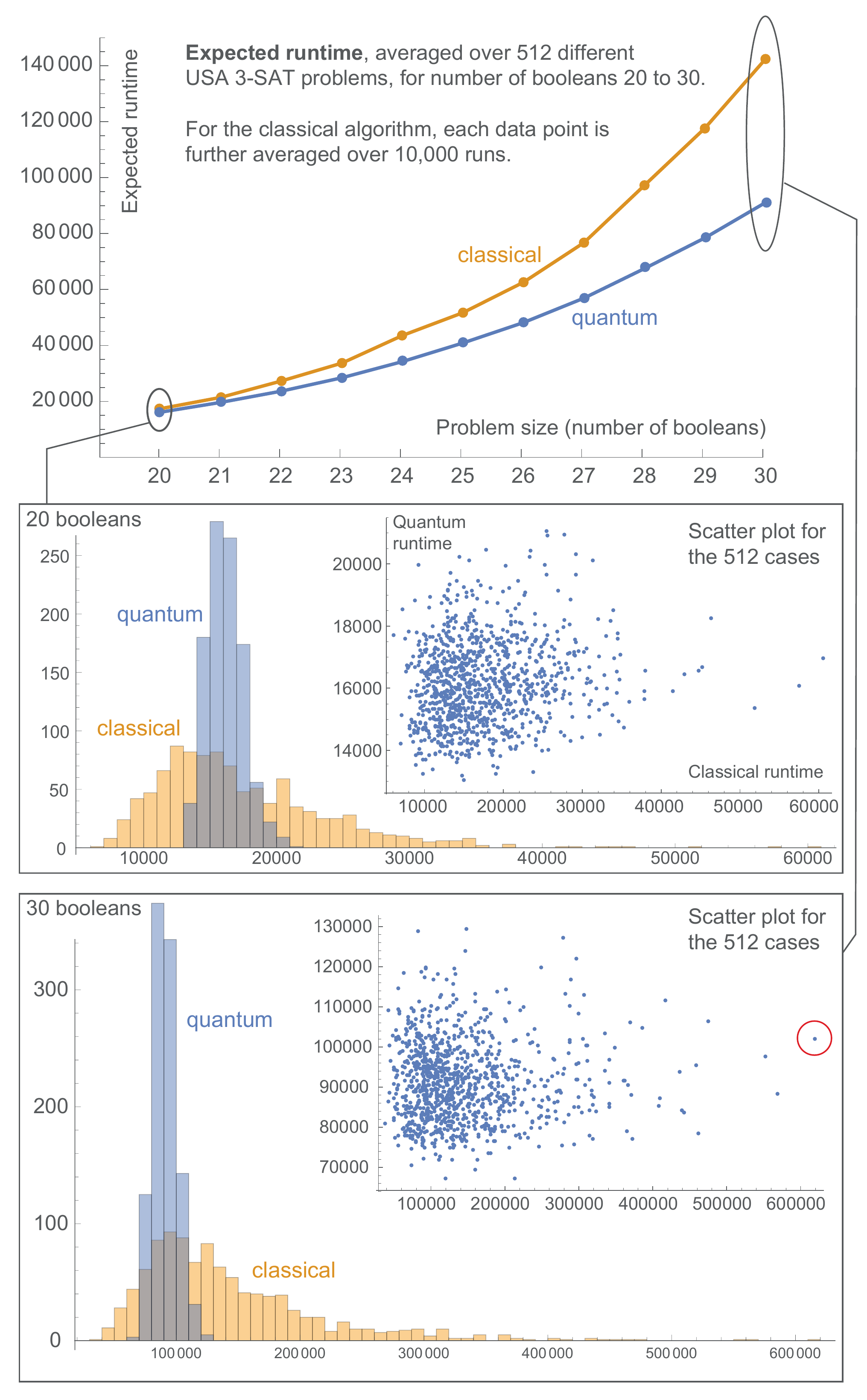}
\caption{A comparison of the expected runtime for the quantum and classical algorithms. The quantum algorithm uses the `evolving $\theta$' approach, similar to Fig.~\ref{fig:34qubits}(b). The initial $\theta$ is set differently according to the number of qubits $n$, ranging from $(0.7)\frac{\pi}{2}$ for $n=20$ to $(0.54)\frac{\pi}{2}$ for $n=30$. The quantum algorithm exhibits a smaller variance in the runtime, and the curves can be fitted quite well by exponential functions (scaling as $1.23^n$ for the classical, $1.19^n$ for the quantum). However this is not conclusive --  such a limited set of data can be fitted even by polynomial functions. For comparison, a na\"ive use of Grover's algorithm would scale as $1.41^n$, while a Grover-accelerated Sch\"oning algorithm might scale as $1.14^n$.  }
\label{fig:classicalComp}
\end{figure}

Having examined the scaling of the characteristic timescale of the quantum problem, it is interesting to make a relative comparison versus the classical algorithm mentioned earlier. Recall that this classical algorithm has a superior asymptotic scaling to Grover's algorithm ($1.334^n$ versus $1.414^n$ for Grover's) so it should be interesting to see if the present quantum algorithm can be competitive. For the purposes of this comparison we adopt the `evolving-$\theta$' approach exemplified by the right-hand panel of Fig.~\ref{fig:34qubits}. We will evaluate the number of individual clause check events that must be performed before finding the satisfying solution to the {\small 3-SAT}, since both the quantum and the classical algorithm require repeated clause checks. The expected number of clause check operations required will be referred to as the `runtime'. The classical algorithm is a random search and so its runtime for a given problem varies each time we run it; the numbers shown presently are an average over $10,000$ runs.

Quantifying the ideal performance of the quantum approach would involve finding the best way to vary $\theta$  with the cycle number $c$ (i.e. the blue curve in the lower panel of Fig.~\ref{fig:34qubits}). However, given the computational costs we simply adopt the cubic form given in Eqn.~(\ref{eqn:thetaEvolve}) for all cases.
There are still two free parameters, $c_Q$ (the total number of clause cycles we attempt) and $\theta_\text{init}$ (the initial value of evolving parameter $\theta$). We have roughly optimised these.

For both the quantum and the classical algorithms, the runtime varies with the particular {\small 3-SAT} problem even when those problems contain the same number of booleans (recall that all {\small 3-SATS} have a unique satisfying solution). In Fig.~\ref{fig:classicalComp} we see that the classical algorithm has a far wider range of runtimes than the quantum algorithm. We also notice that the quantum algorithm's average performance is superior to the classical algorithm, and the margin increases with the number of booleans. A third interesting observation is that there is very little correlation between `classically harder' problems for which the classical algorithm needs the longest runtime, and degree to which those particular problems are hard for the quantum algorithm. The instance highlighted by a red circle in the lowest panel of Fig.~\ref{fig:classicalComp} is, for the classical algorithm, the single hardest problem of all those considered; for the quantum algorithm it is merely a typically case. This is consistent with the idea that the quantum process is a fundamentally different kind of solution-finding protocol. 

The largest simulation reported here is for $34$ booleans and therefore $34$ qubits (Fig.~\ref{fig:34qubits}). For the systematic investigation presented in Fig.~\ref{fig:classicalComp}, $512$ different {\small 3-SAT} problems are considered for each problem size, and the size ranges from $20$ to just $30$ qubits. Given only this  limited range, which is determined by the considerable time and memory requirements for exact simulations, one cannot definitely state that the quantum outperforms the classical. In any case one cannot establish the asymptotic scaling of the runtime from numerics. 

Notwithstanding the limitations of the current study, we would say that the simulations are encouraging and the algorithm may represent an interesting and novel approach to seeking satisfying solutions for {\small 3-SAT} problems. It is possible that further work will elucidate the scaling properties; for now we can say that over the range $20-30$ booleans it appears far superior to a direct application of Grover's search, and superior to Sch\"oning's well-known classical algorithm (with a widening margin as the problem size increases). In fairness we should remark that the merit of Sch\"oning's algorithm is that it has provable asymptotic scaling, and in real world applications one might use other classical approaches whose performance is not formally bounded but which may be found to be superior in practice. However the quantum algorithm is also not optimal and many variants might be considered.

\section{Impact of errors}

Finally we remark on degree to which the quantum algorithm is  error resilient, where by errors we encompass both environmental decoherence of the qubits and, more importantly, any imperfections to the gate operations. The first remark one can make is encouraging: given an error rate that is small compared to $1/n$, where $n$ is the number of booleans, we would not expect that errors cause us to reach a `wrong' solution to the {\small 3-SAT}. Each successive clause check cycle in our algorithm effectively suppresses any errors in previous cycles: either the register is projected back towards the `correct' state, or a clause check operator will fail and so require a restart. Thus if we are able to reach the end of our quantum algorithm, having performed the desired $c_Q$ clause check cycles, it is likely that the resulting state is properly correlated to the classical solution (and if we have evolved $\theta$ to $\theta=\tfrac{\pi}{2}$ then it {\it is} the classical solution). 

However, the cost of a finite error rate will manifest itself in the increased probability of our algorithm failing before we reach $c=c_Q$ complete clause cycles, thus necessitating more restarts. Therefore the required runtime will increase. To take a specific example, consider a depolarising error suffered by one of the qubits, when the state of the register is close to the desired $\ket{\bm \theta}$. We can regard the depolarising error as a Pauli $X$, $Y$ or $Z$ flip selected uniformly at random and applied to the qubit. Equivalently we can select from these same three operations in a rotated basis; if identify the $Z$-eigenstates with the basis defined by $\{\ket{\theta},\ket{\theta_\perp}\}$ where $\ket{\theta}$ is the correct state of the qubit, then we see that a $Z$ error is harmless whereas an $X$ or $Y$ flip will leave the qubit in state $\ket{\theta_\perp}$, with the consequence that a clause check operator must fail in the next cycle. Thus if our error model calls for us to apply a Pauli error with some probability $p_\text{error}$ the consequence is that the algorithm will imminently abort with probability $\tfrac{2}{3}p_\text{error}$. Consequently our probability of successfully reaching $c=c_Q$ cycles will fall exponentially with $n_gp_\text{error}$ where $n_g$ is the number of gates, which in turn scales at least linearly with problem size. We conclude that the algorithm is not error-resilient in any strong sense.

While our algorithm may not resist errors, it might be that it is compatible with relatively simple error correction techniques. A novel feature is that the desired state $\ket{\bm \theta}$ is a product state of the qubits, so that any encoding/monitoring process need not preserve entanglement between qubits, only the integrity of individual qubits. 

\section{Conclusion}
In conclusion, we have presented a quantum algorithm which seeks the solution to {\small 3-SAT} problems through a filtering process. We apply clause check operations in a basis according to which the quantum states for {\small TRUE} and {\small FALSE} are not orthogonal. The parameter  $\theta$ defining the degree of orthogonality can be fixed, or may be allowed to evolve in an adiabatic-like fashion. In either case the algorithm is very simple, and yet compares favourably with a classical algorithm that has one of the best known asymptotic performances. 

\bigskip
\noindent{\bf Acknowledgements:} Our thanks to Wim van Dam, Earl Campbell, Tyson Jones and Ashley Montanaro for helpful conversations.  JFF acknowledges support from the Air Force Office of Scientific Research under AOARD grant no. FA2386-15-1-4082. This material is based on research supported in part by the Singapore National Research Foundation under NRF Award No. NRF-NRFF2013-01. SCB acknowledges support from the Engineering and Physical Sciences Research Council National Quantum Technology Hub in Networked Quantum Information Technologies.


\bigskip

\appendix

\section{{\small 3-SAT} problems generated here}
\label{appendix:3SATgeneration}

For the numerical simulations described presently, it was necessary to generate large numbers of {\small 3-SAT} problems. They were generated  randomly but in such a way that the following rules applied:

\begin{enumerate}
  \item \label{unique} All clauses are distinct from one another.
  \item \label{TandF} Every variable $b_i$ occurs at least once in the {\small 3-SAT} proposition in positive form $b_i$, and at least once in negated form $\lnot b_i$.
  \item \label{diffVar} All clauses  involve three different variables, thus neither $(b_1\lor b_1 \lor b_2)$ nor $(b_1\lor \lnot b_1 \lor b_2)$ could be present.
\item \label{5times} Given $n$ variables we generate $m={\rm round}(R\ n)$ clauses, with $R=4.267$.
\end{enumerate}

Having generated a {\small 3-SAT} formula in this way we then determine the number of solutions, $n_S$. This is a simple task classically for the problem sizes we consider. Generally we will wish to have {\small 3-SAT} problems with a specific $n_S$ for testing the quantum algorithm; if the randomly generated {\small 3-SAT} has some other $n_S$ we discard it and generate a fresh formula. For the majority of data we present here we have chosen $n_S=1$, i.e. problems with a `unique satisfying assignment' (USA), however we do also consider cases with $n_S>1$. While it is quite rare to have exactly one solution, so that these {\small 3-SAT} instances are special, we choose this to reduce variably in the challenge of solving them. Note that criterion  (\ref{unique}) is merely to ensure proper counting of the number of clauses -- if two clauses were identical, one could be dropped without changing the {\small 3-SAT} problem. Similarly (\ref{TandF}) is imposed because otherwise the effective number of variables is reduced: for example if a variable only occurs in positive form $b_i$ without $\lnot b_i$ appearing anywhere in the formula then trivially one chooses $b_i$={\rm\small TRUE} and all clauses involving $b_i$ become ${\rm\small TRUE}$. Meanwhile (\ref{diffVar}) is merely for simplicity -- in principle recurrence of a variable within a clause is legitimate for {\small 3-SAT} but we do not consider such cases (although there is no apparent difficulty with such a generalisation). Point (\ref{5times}) is to maximise the chances that a randomly generated set of clauses will have a small but non-zero number of solutions $n_S$. We rely on the results of previous SAT studies which have found that there is a key threshold in the ratio $R$ between the number of clauses and the number of variables -- instances below the threshold are likely to be satisfied with multiple solutions, and instances above it are unlikely to be satisfiable. This threshold has not been absolutely determined, and it only becomes sharp as $n\rightarrow\infty$, but the value $4.267$ has been suggested~\cite{nature4p267}.

\bigskip

\section{Fidelity is monotonic}
\label{appendix:FidelityMonotonic}

\smallskip

This section considers the evolution of the qubit register's fidelity with respect to the target state, for a {\small 3-SAT} with a unique satisfying argument. However the extension to the multi-solution fidelity (next section) is straightforward. 

As observed by van Dam~\cite{vDam02}, for certain adiabatic algorithms one can construct {\small 3-SAT} problems for which the evolution `gets stuck' and never finds the actual solution. Generally this is a potential issue in any approach (quantum or classical) that seeks a solution by minimising a cost function. A simple example of a cost function is the number of unsatisfied {\small 3-SAT} clauses. An approach that involves updating the system's state so as to follow the cost function `down hill' to reach the lowest possible cost, may run the risk of being trapped in a local minimum (and therefore will require a strategy for escaping such minima). 

The present approach does not appear to have this issue: The fidelity of the qubit register with respect to the target state will only increase with each complete cycle of the clause checks. (This is not unexpected since the obvious quantum approach, i.e. Grover's algorithm, also does not involve a cost function and cannot become trapped.) Generally we can write the state of our register as 
\[
\ket{\Psi}=\alpha\ket{\bm\theta}+\beta\ket{\bm\theta^\bot}
\]
where ${\bm\theta}$ is the desired state as ${\bm\theta^\bot}$ is some state orthogonal to the desired state. The fidelity is $|\alpha|^2$ and providing that this is non-zero then when the next clause check operation is performed there is a finite probability of `passing' that check; then the post-check state is 
\[
\ket{\Psi}=\mathcal{N}(\alpha\ket{\bm\theta}+\beta^\prime\ket{\bm\theta^\bot})
\]
where $\beta^\prime\leq\beta$, normalisation constant $\mathcal{N}\geq 1$. 

The inequalities are not strict because we have not excluded the possibility that the three qubits involved were certain to pass the clause check -- they could have been in a superposition of several states, all of which satisfy the clause in question but only one of which satisfies the {\small 3-SAT}. However a {\it complete cycle} of successful clause check operations must indeed reduce $\beta$, unless it was already zero, because (as discussed earlier) $\ket{\bm\theta}$ is precisely the {\it unique} state that passes all clause checks with certainty.

Of course this leaves open the question of whether one can adversarially construct a {\small 3-SAT} problem that has an extremely slow evolution toward the target state $\ket{\bm \theta}$.
 
\section{Tracking $c_\text{smooth}$}
\label{appendix:cSmooth}

\begin{figure}[!b]
\centering
\includegraphics[width=0.8\linewidth]{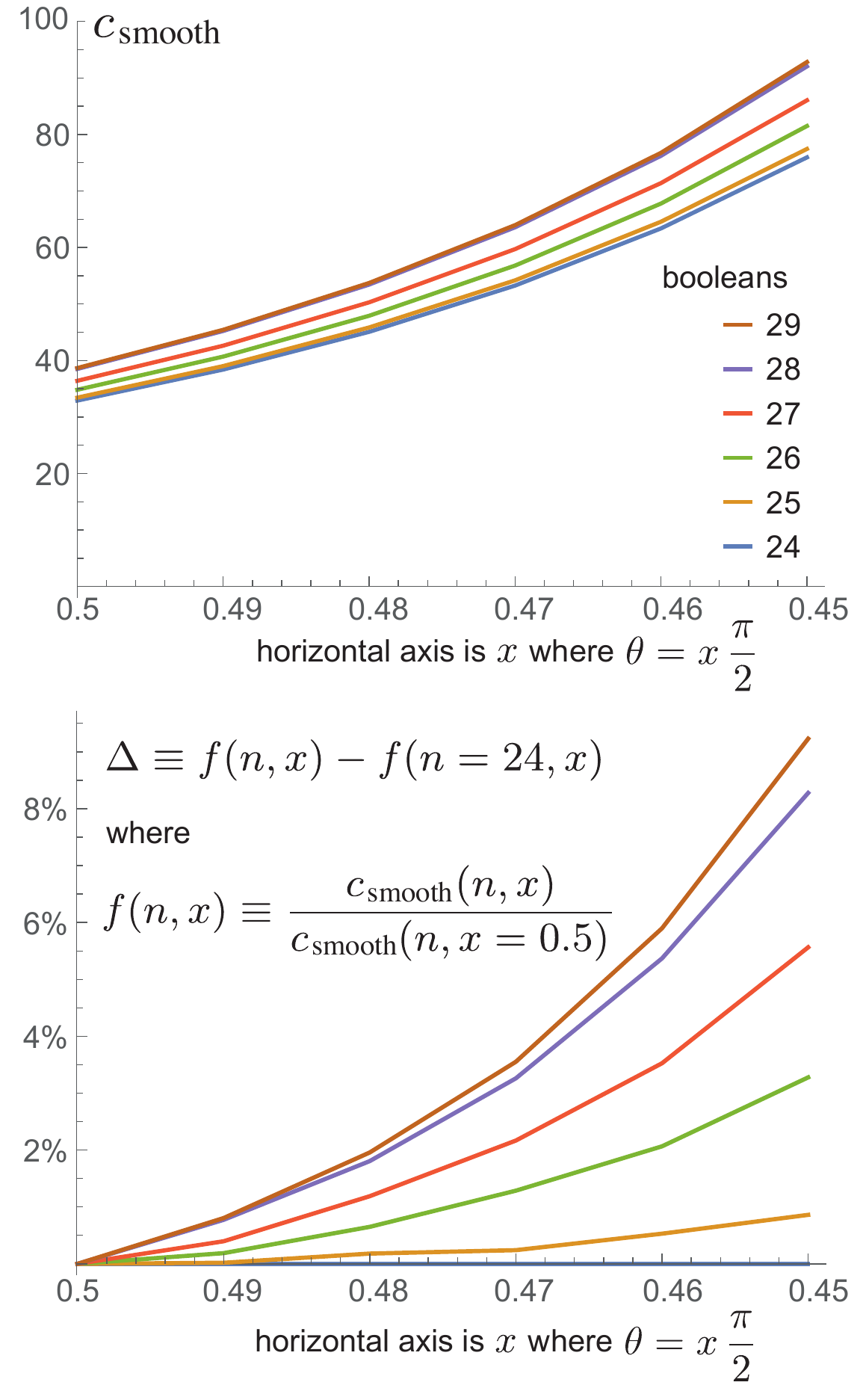}
\caption{Upper plot shows how the quantity $c_\text{smooth}$ evolves as we vary the problem size (between $24$ and $29$ booleans) and we reduce the internal parameter $\theta$. Lower plot shows $\Delta$, a scaled and shifted quantity emphasising the relative curve shapes. Each data point is derived from $512$ random \small{3-SAT} problems created as explained in Appendix~\ref{appendix:3SATgeneration}. Notice that this is sufficient sampling to reveal the broad trends, i.e. more booleans implies a higher $c_\text{smooth}$  on average, but it is insufficient sampling to establish the functional dependence (the vertical spacing of the lines is irregular). }
\label{fig:nSmooth}
\end{figure}

The quantity $c_\text{smooth}$ is defined in the main text, it is the number of complete clause cycles which much be successfully completed in order that all qubits have a probability of at least $0.51$ of being found with the `correct' alignment when measured in the $z$-basis. The results are shown in Fig.~\ref{fig:nSmooth}.

\section{Information from measuring a $\theta<\frac{\pi}{2}$ state}
\label{appendix:InfoFromThetaBelowMax}

\smallskip

Suppose that a high fidelity approximation to the state $\ket{\bm \theta}$ with $\theta<\frac{\pi}{2}$ has been created by e.g. the fixed-theta algorithm. This state is simply a product of individual qubit states $R_Y(\pm \theta)\ket{+}$ where the sign is positive if the corresponding boolean is {\small FALSE} in the unique satisfying assignment (USA), and negative otherwise. If a given qubit is measured in the z-basis then the measurement will give the correct truth value (according to rule $\ket{0}\rightarrow$~{\small FALSE} and $\ket{1}\rightarrow$~{\small TRUE}) with a probability $p=(1+\sin\theta)/2$. In the following we will call a measurement `incorrect' if it implies the wrong boolean value; note that we do not mean to suggest some failure in the measurement apparatus -- these we are taking to be perfect.  Measurements are independent in the sense that the question of whether a given qubit  gives the correct value is not correlated to the correctness of any another qubit measurement. (Note that here we are considering the exact state $\ket{\bm\theta}$, whereas of course the real state created using e.g. the fixed-theta protocols will not be exact and may include highly correlated elements -- but for a sufficiently high fidelity the remarks here will be valid.)

Having measured all qubits and noted the results, we can repeat the process of creating $\ket{\bm \theta}$ and again measure it. This new set of measurements will be independent of the former set. So the probability of a measurement being correct is $p$ for every qubit and on every run. (Notice that one might adopt a superior algorithm where the initial state is not $\ket{+}^{\otimes n}$ but is instead adapted to our current `best guess' of each qubit's solution state orientation -- but the present analysis assumes we just na\"ively keep repeat exactly the same process).

If we create and then measure out $k$ states, each close to $\ket{\bm \theta}$, then our best guess as to the {\small 3-SAT}'s solution is to assign each boolean according to the majority vote from the $k$ measurements of the corresponding qubit. The Chernoff bound then implies that the probability of guessing a given boolean incorrectly falls as some exponential function $\exp(-Ck)$, and thus the probability of guessing {\it any} of the $n$ booleans incorrectly is bounded by $n\exp(-Ck)$. Thus the number of states one must create and consume before finding the classical solution is only logarithmic in $n$. 

\smallskip 
\noindent{\bf An explicit analysis}

The scenario with $n$ booleans is equivalent  to the following classical problem: we are given $n$ numbered coins, each with the same known bias $p>\frac{1}{2}$, but we do not know whether a given coin favours `heads' or `tails'. We are allowed to throw all coins at once and note the outcomes; how many such throws do we need before we can guess {\it all} the biases correctly?

There are several ways to make this question precise and then derive an answer. We prefer the following: What is the expected number of throws, $R$, required until our {\it best guess} of the set of boolean values is {\it probably error free} -- i.e. the guess is fully correct with probability greater than a half. Now after $R$ throws we make our best guess about the bias of a given coin, i.e. head-favouring or tail-favouring, simply by noting whether it produced more heads or more tails (we can assume $R$ is odd to avoid even-splits where we don't know how to guess). Let's say a given coin is head-favouring. Then the probability that we wrongly guess `tail-favouring!' is given by the sum of the probabilities that the number of heads, $i$, is less than half of $R$.
\[
p_{\rm wrong}=\sum_{i=0}^{i=M} \left(\begin{array}{c}R \\i\end{array}\right) p^i(1-p)^{R-i},
\]
where $M=(R-1)/2$. 
In order to achieve a good probability that no such mistake will occur over the entire set of $n$ coins, we can require that $p_{\rm wrong}\lesssim n^{-1}$ assuming that $n$ is large, i.e. 
\begin{equation}
\sum_{i=0}^{i=M} \left(\begin{array}{c}R \\i\end{array}\right) p^i(1-p)^{R-i}<\frac{1}{n}.\nonumber
\end{equation}

If the product $R(1-p)$ is greater than about $5$, the sum is well approximated by an integral. Making that assumption, we have:
\begin{equation}
\frac{1}{\sigma\sqrt{2\pi}}\int_{-\infty}^{R/2} \exp\left(-\frac{(x-Rp)^2}{2\sigma^2}\right)\ dx<\frac{1}{n} \nonumber
\end{equation}
where the variance $\sigma^2=Rp(1-p)$. Rescaling $x$ yields
\[
\frac{1}{\sqrt{\pi}}\int_{-\infty}^{-G} \exp(-x^2)\ dx<\frac{1}{n}\ \ \ {\rm where}\ \ \ G=\frac{\left(p-\frac{1}{2}\right)\sqrt{R}}{\sqrt{2p(1-p)}}.
\]
We recognise the complimentary error function erfc,
\[
\frac{1}{2}{\rm erfc}(G)<\frac{1}{n}
\]
and recall that to lowest order (for large $x$)
\[
{\rm erfc}(x)\simeq\frac{\exp(-x^2)}{x\sqrt{\pi}}
\]
so that the criterion becomes
\[
G\exp(G^2)>\frac{n}{2\sqrt{\pi}}.
\]
Now returning from coins to qubits and noting the assumptions made, we conclude: When attempting to solve USA {\small 3-SAT} problems of $n$ booleans using the fixed-theta algorithm with $R$ repetitions of the ``prepare $\ket{\bm \theta}$ and measure'' cycle, $R$ {\it need only increase logarithmically} with problem size $n$. 

Moreover we recall that bias $p=(1+\sin\theta)/2$ so we can rewrite 
\[
G=\frac{\sin\theta\sqrt{R}}{\sqrt{2(1+\sin\theta)(1-\sin\theta)}}=\sqrt\frac{R}{2}\tan\theta.
\]

We might also consider a {\small 3-SAT} of a fixed size and ask how the selected value of $\theta$ influences the expected number of repetitions $R$ needed to guess all booleans correctly. This can be assessed simply by requiring $G$ to stay constant, so 
$R\propto\cot^2\theta$ and thus $R\sim\theta^{-2}$ for small $\theta$.

\section{Why not do a full adiabatic sweep?}
\label{appendix:whyNotFullSweep}

\begin{figure}[t]
\centering
\includegraphics[width=.9\linewidth]{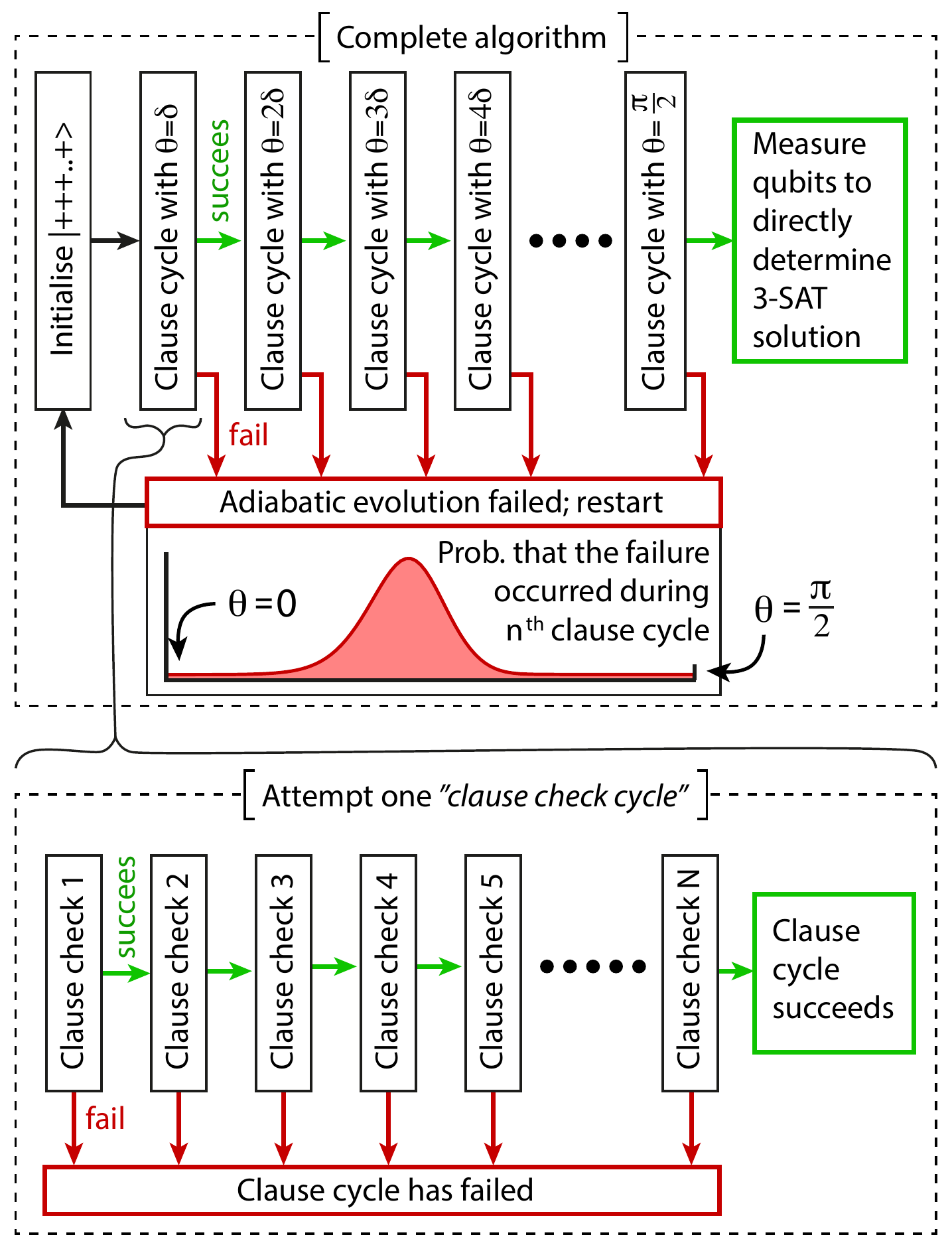}
\caption{ Flow diagram for the basic `adiabatic-like' algorithm, starting from $\theta=0$. This approach is inefficient.
\label{fig:alg_flow}
}
\end{figure}

In the main paper, the quantum algorithm was presented as starting with some $\theta$ value $\theta_\text{init}$ that is non-zero. This value might then remain constant or might increase until $\theta=\tfrac{\pi}{2}$. For the latter case where we are going to allow the $\theta$ value to evolve, the reader might wonder why we don't just start from $\theta=0$. Such a choice is possible, but will not be optimal. The reason is that the maximum chance of the algorithm failing occurs mid-way through the process, thus wasting all the time invested in reaching that mid-way point. This is shown in Fig.~\ref{fig:alg_flow}.

\begin{figure}
\centering
\includegraphics[width=.75\linewidth]{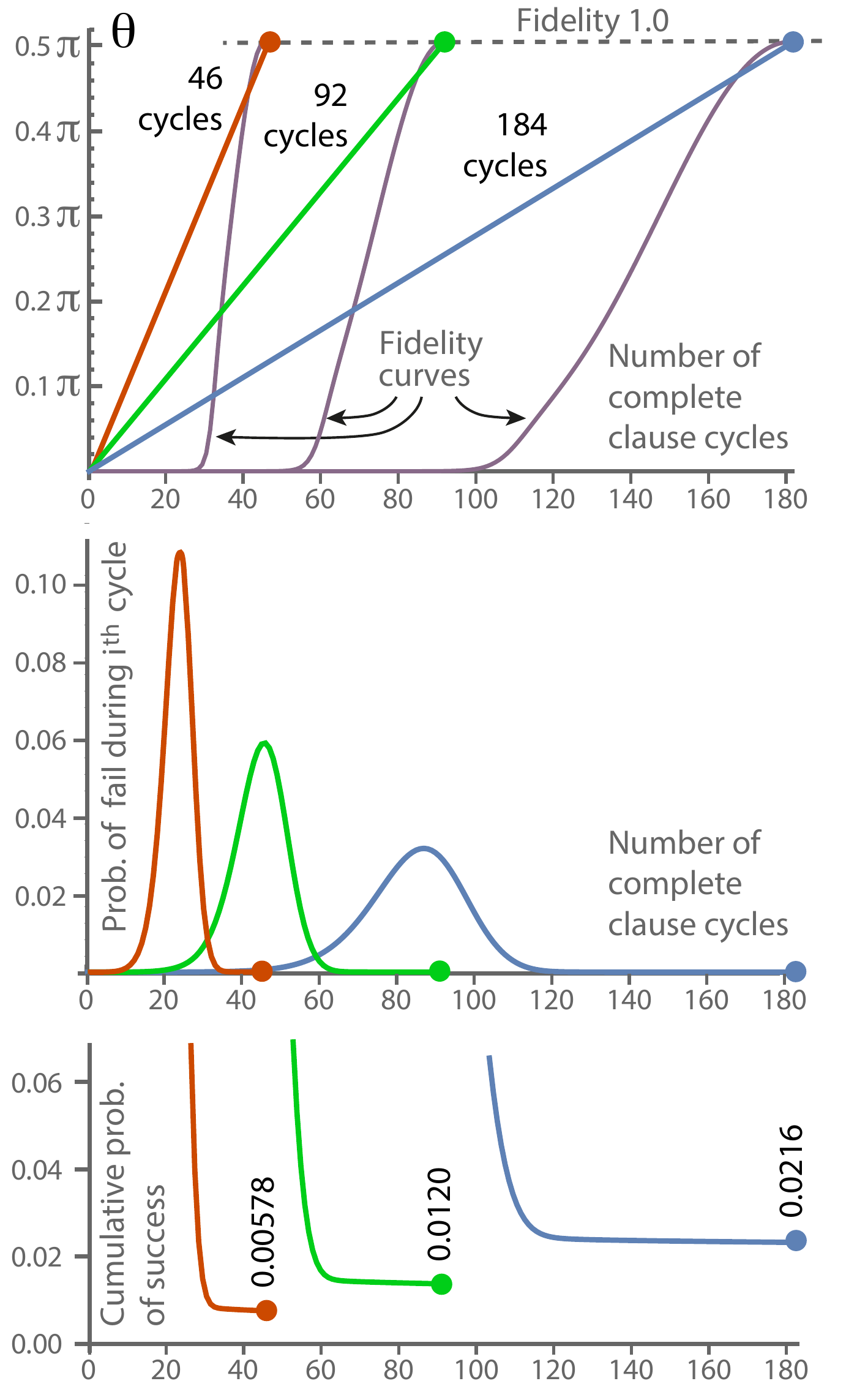}
\caption{Data from a simulation of the `adiabatic-like' approach (Fig.~\ref{fig:alg_flow}) with a {\small 3-SAT} problem of $24$ booleans and a unique solution. The red, green and blue traces correspond to three different rates of {\it linear} evolution of the parameter $\theta$. The slower the evolution, the higher the chance of reaching the desired state ultimate state $\ket{{\bm\theta}=\frac{\pi}{2}}$; the fidelity with respect to this target is shown by the grey curves in the uppermost figure. However it is {\it not} optimal to go very slowly, as explained in the text.}
\label{fig:simpleLinearPart1}
\end{figure}

One might imagine that it is best to make $\delta$, the per-cycle increment to $\theta$, as small as is necessary to make the probability of success approach unity. This intuition proves to be incorrect. To understand why, it is instructive to look at a simulation of a particular problem; in Fig.~\ref{fig:simpleLinearPart1} we show data for a typical USA {\small 3-SAT} problem involving $24$ booleans. We see that, as expected, a slower evolution leads to a better chance of reaching the desired $\theta=\frac{\pi}{2}$ final state.

However, in fact one should not evolve more slowly than a certain optimum, and this optimum corresponds to a success probability that is still far below unity. This is because there is a peak to the failure probability when $\theta$ is already substantial, around in the mid-point in the evolution as can be seen from the middle graph of Fig.~\ref{fig:simpleLinearPart1}. The lowest graph in Fig.~\ref{fig:simpleLinearPart1} shows that doubling the number of increments from $92$ (green) to $184$ (blue) causes the success probability for a given `run' to increase. However the expected number of steps-until-failure on a each failed run {\it increases by a greater factor} because the peak probability of failure is now occurring after about twice as many cycles. The net consequence is a slight {\em increase} in overall expected number of clause checks required before a {\small 3-SAT} solution is found. In this particular problem the green curve is actually the optimal choice among the three curves. However the {\small 3-SAT} would be solved far more efficiently by starting from a substantial $\theta$ value, as in the algorithms considered in the main paper.

\section{Proof that $n_S$ classical solutions implies $n_S$ quantum solutions}
\label{appendix:NoOtherSolutions}

\smallskip

In the main paper, we introduced a state in Eqn.~\ref{eqn:solutionStateTheta} i.e.
\[
\ket{\bm \theta}=\bigotimes_i \ R_Y\left(L_i\theta\right)\ket{+}_i 
\]
with $L_i=+1$ or $-1$ according to the truth value of boolean $i$ in the unique satisfying solution to the classical \small{3-SAT}. This state has the property that it will `pass' all clause check operators, i.e. 
\[
\mathcal{C}_i\ket{\bm \theta}=\ket{\bm \theta}\ \ \forall i.
\]
Consequently, this state will not evolve when the clause checks are applied; it is a final fixed-point in that sense. We will refer to this fixed-point as a {\it quantum solution}. If the classical \small{3-SAT} problem has exactly one solution, we wish to know whether there is also exactly one quantum solution. More generally, if there are $n_S$ satisfying solutions for a classical \small{3-SAT} (where $n_S$ may be zero), will there be exactly $n_S$ independent quantum solutions -- i.e. an $n_S$-dimensional subspace of solutions? 

The answer is yes. Before attempting to prove this, it is convenient to review and slightly extend the earlier notation.

\smallskip
\smallskip

{\bf \noindent Notation}

Following the main paper we define
\[
\ket{\theta^i}=R_Y\left(+\theta\right)\ket{+}_i\ \ \ \ {\rm and}\ \ \ \ \ket{{\bar \theta}^i}=R_Y\left(-\theta\right)\ket{+}_i.
\] 
The use of the superscript $i$ always implies that this is a state of the $i^\text{th}$ qubit. Notice that if, say, the classical \small{3-SAT} has a unique solution involving $b_i\rightarrow$\small{TRUE} then by this definition $\ket{\theta^i}$ is the `correct' state of qubit $i$, whereas $\ket{{\bar \theta}^i}$ is the `incorrect' state; it {\it would be} correct if boolean $b_i\rightarrow$\small{FALSE} in the satisfying classical solution. We will also introduced
\[
\ket{\theta^i_\perp}=R_Y\left(\pi+\theta\right)\ket{+}_i\ \ \ \ {\rm and}\ \ \ \ \ket{{\bar \theta_\perp}^i}=R_Y\left(\pi-\theta\right)\ket{+}_i.
\] 
These are the states which are orthogonal to $\ket{\theta^i}$ and $\ket{{\bar \theta}^i}$ respectively, i.e. 
\[
\braket{\theta^i_\perp}{\theta^i}=0\ \ \ \ \ {\rm and}\ \ \ \braket{{\bar \theta}^i_\perp}{{\bar \theta}^i}=0.
\]
As noted earlier the states $\ket{\theta^i}$ and $\ket{\bar\theta^i}$ are not orthogonal, since we are considering $0<\theta<\frac{\pi}{2}$ (note the inequalities are strict). In fact we noted that
\[
\braket{\theta^i}{\bar\theta^i}\neq 0,\ \ \ \braket{\theta^i_\perp}{\bar\theta^i_\perp}\neq 0,\ \ \ \braket{\theta^i_\perp}{\bar\theta^i}\neq 0,\ \ \ \braket{\bar\theta^i_\perp}{\theta^i}\neq 0.\ \ \ 
\]
For each of the clause check operators $\mathcal{C}_i$ we can write
\[
\mathcal{C}_i\equiv\mathcal{I}-\mathcal{P}_i
\]
where $\mathcal{I}$ is the identity and projector $\mathcal{P}_i$ was defined by Eqn.~(\ref{eqn:CiDef}), which we extend slightly as
\[
\mathcal{P}_i\equiv \ket{A}\bra{A}\ \ \ {\rm or}\ \ \ket{AB}\bra{AB}\ \ \ {\rm or}\ \  \ket{ABC}\bra{ABC}
\]
so that we now permit the clause to involve one, two or three booleans (rather than necessarily three). Here $\ket{A}$, $\ket{B}$ and $\ket{C}$ each stand for a single qubit state, either $\ket{\theta_\perp}$ or $\ket{{\bar \theta_\perp}}$ according to what is `wanted' by the clause, just as explained in the preamble before Eqn.~(\ref{eqn:CiDef}).

We say that a state $\ket{s}$ is a quantum solution if satisfies or `passes' all the clause check operators:
\begin{equation}
{\mathcal C}_i\ket{s}=\ket{s}\ \ \forall i\ \ \ \ \Leftrightarrow\ \ \ \ {\mathcal P}_i\ket{s}=0\ \ \ \forall i.
\label{eqn:basicProj}
\end{equation}

\bigskip

{\bf \noindent Proof} 

\begin{figure}[b]
\centering
\includegraphics[width=0.9\linewidth]{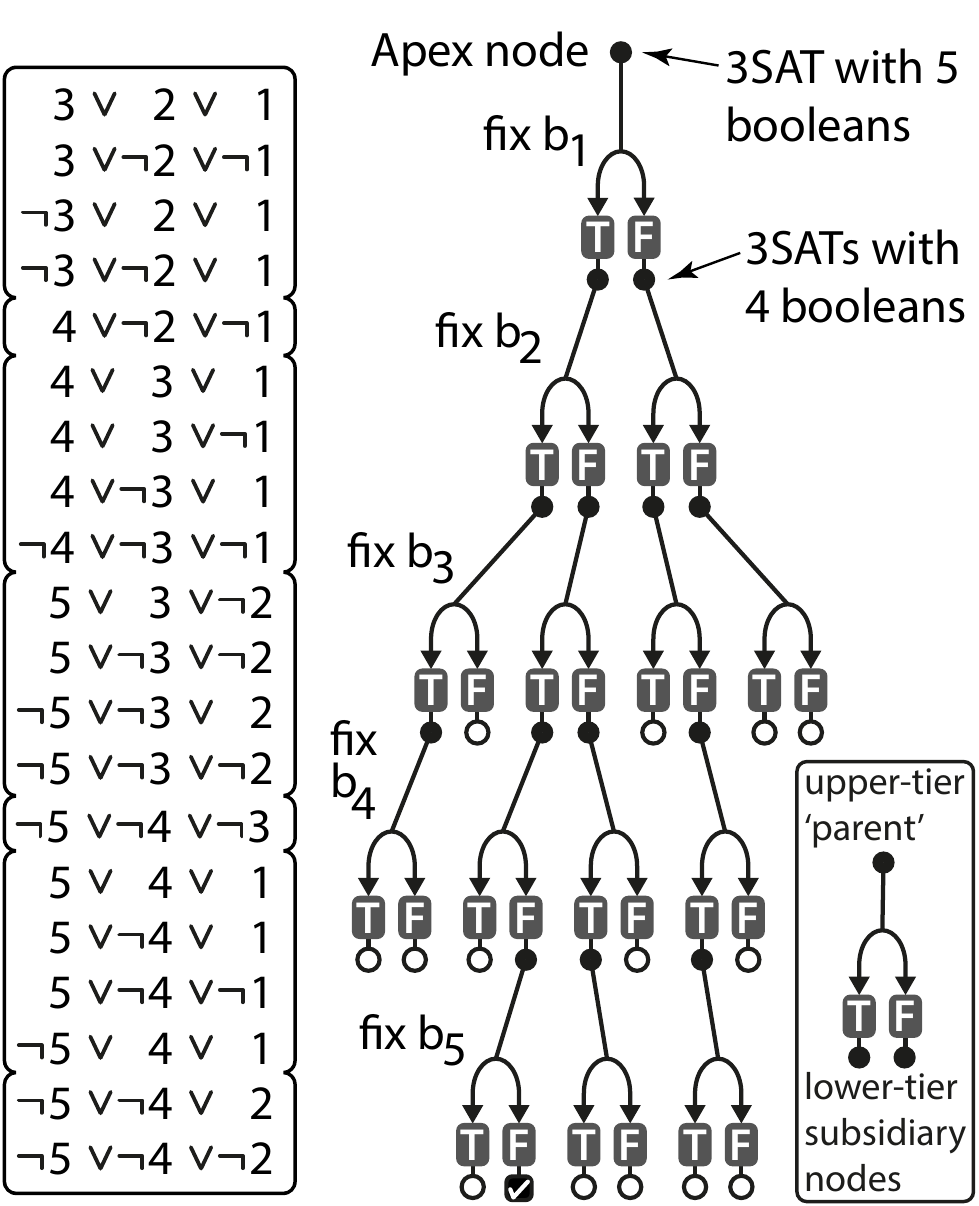}
\caption{ Schematic showing relationship between classical \small{3-SAT} problem of five boolean variables $b_1$ to $b_5$. Left panel shows the full set of clauses in standard notation. In the root structure, filled circles are \small{3-SAT} problems while empty circles are propositions that have evaluated to \small{FALSE} and the tick-in-square symbol is a proposition that has evaluated to \small{TRUE}, i.e. a successful solution to the apex \small{3-SAT}. 
}
\label{fig:3sat_graph}
\end{figure}

We now introduce the tree structure shown in Fig.\,\ref{fig:3sat_graph}. The structure relates purely to classical \small{3-SAT} problems. At the apex of the tree is a node representing a full \small{3-SAT} problem of $n$ variables, where $n=5$ in the illustrated example. We then construct a root structure below the apex, by taking each boolean in turn and setting it to either  \small{TRUE} or  \small{FALSE}. Each binary branching yields two subsidiary  \small{3-SAT} problems, simply by substituting the  \small{TRUE} or  \small{FALSE} value for the boolean which has been fixed, and simplifying the \small{3-SAT} thus:
\begin{enumerate}
\item  \label{case1} If any given clause, taken in isolation, now evaluates to \small{FALSE} then the entire \small{3-SAT} proposition evaluates to \small{FALSE}. This terminates the particular root path, and is denoted by an open circle in the figure.
\item \label{case2} Conversely if {\it every} clause, taken in isolation, now evaluate to \small{TRUE}, then we have a satisfying solution to the \small{3-SAT}, denoted by a square with a tick symbol.
\item Otherwise, we delete any clause that now evaluates to \small{TRUE}, and simplify any clauses that are now of the form $(\text{\small FALSE}\lor X \lor Y)$ or $(\text{\small FALSE}\lor Z)$ to $(X \lor Y)$ or $(Z)$ respectively.  This yields a new \small{3-SAT} proposition that is not manifestly either \small{TRUE} or \small{FALSE}, and we mark this new node of the structure as a filled circle.
\end{enumerate}

\begin{figure}[b]
\centering
\includegraphics[width=0.9\linewidth]{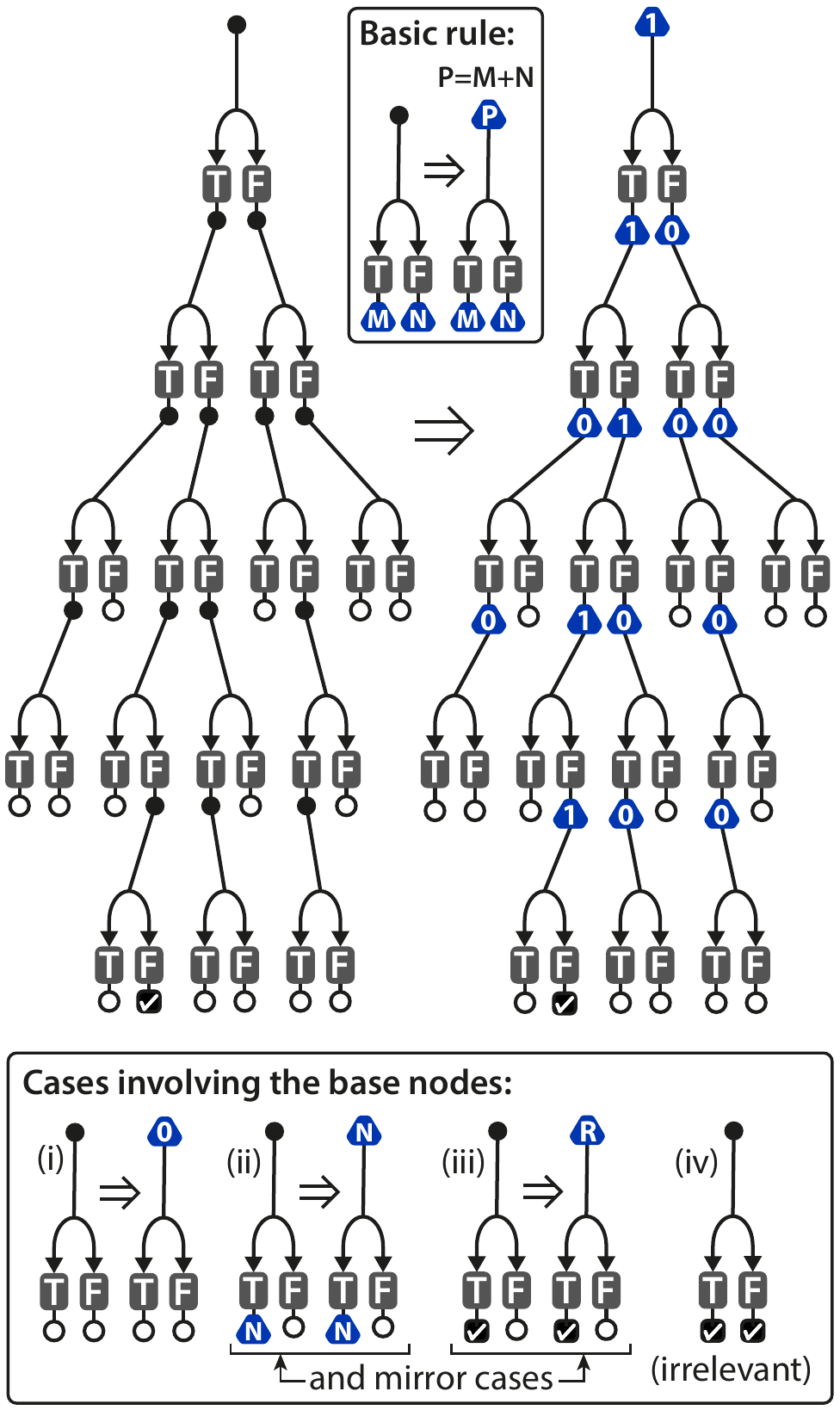}
\caption{ Schematic showing how we can infer the number of quantum solutions for any filled-circle node. The numbers in the blue triangles indicate how many independent solutions exist for the quantum problem corresponding to a given \small{3-SAT}. Note the in the lower panel part (iii), $R=2^q$ where $q$ is the number of unfixed booleans in the subsidiary problems.  
}
\label{fig:3sat_quantumCase}
\end{figure}

Having established this hierarchy of classical \small{3-SAT} problems, we now address the question of how many quantum solutions exist for the analogous quantum problems. We answer this by focusing on an arbitrary node within the structure, and its two subsidary nodes; initially we will limit ourselves to the case that both subsidiary nodes are filled circles: their \small{3-SAT}s are not manifestly \small{TRUE} or \small{FALSE} and indeed they contain no clause which, regarded in isolation, is either \small{TRUE} or \small{FALSE}. We may use the term `parent' for the upper node, see right-inset in Fig.\,\ref{fig:3sat_graph}. We will show if the subsidiary nodes have $M$ and $N$ quantum solutions, respectively, where $M$ and $N$ are any integers or zero, then the parent node has $M+N$ solutions. This applies when neither of the subsidiary classical \small{3-SAT}s resolve directly to \small{TRUE} or \small{FALSE} in the sense of steps (\ref{case1}) and  (\ref{case2}) above; presently we will consider these as special cases. These rules allow us to infer the number of quantum solutions to the base levels of the root structure, and from them to infer the number of quantum solutions for every node. This is illustrated in Fig.~\ref{fig:3sat_quantumCase}. We notice that for the apex \small{3-SAT}, and indeed every subsidiary \small{3-SAT}, the number of quantum solutions matches the number of classical solutions.

Without loss of generality, suppose that the upper `parent' node has boolean variables $b_k$, $b_{k+1}$,...,$b_N$, so that the two lower {\it subsidiary} nodes have the same booleans except $b_k$ which has been set either to \small{TRUE} or \small{FALSE}. 

We will consider the set $\{\mathcal{P}^U\}$ of quantum clause projectors for the upper tier \small{3-SAT}. There are three subsets of $\{\mathcal{P}^U\}$ to examine. The first subset corresponds to those clauses which do not involve our chosen boolean $b_k$. The relevant subset of quantum projectors, call them set $\{\mathcal{X}\}$, will not involve qubit $k$. Any such projector will also apply to the two lower-tier nodes, i.e. the full set of classical \small{3-SAT} clauses associated with either subsidiary node must include a subset for which the corresponding quantum projectors are $\{\mathcal{X}\}$.

Next consider those clauses in the upper-tier \small{3-SAT} which do involve boolean $b_k$, and which `want' this boolean to be \small{TRUE}. We can label the corresponding clause projectors as the subset $\{\mathcal{T}\}$. Now the \small{3-SAT} in the left branch in our root structure, i.e. the subsidiary problem for which indeed we have set $b_k\rightarrow$\small{TRUE}, does not inherit these clauses -- the clauses themselves are all trivially \small{TRUE}. Conversely the right branch leads to the  subsidiary problem where we have set $b_k\rightarrow$\small{FALSE}. This \small{3-SAT} {\it does} inherit the clauses in question, with the simplification that we now remove the reference to $b_k$. Noticing that every projector in  $\{\mathcal{T}\}$ will involve $\ket{\theta_\perp^k}\bra{\theta_\perp^k}$, we can generate the projectors for the subsidiary node thus: 
\[
\{\mathcal{T}^\prime\}\ \ \ \text{such that}\ \ \ \ \mathcal{T}_i^\prime\equiv\bra{{\theta}^k_\perp}\ \mathcal{T}_i\ \ket{{\theta}^k_\perp}.
\]
For the cases we are presently considering, where both subsidiary nodes are filled circles, it is not possible that the parent node has among its clauses the simple clause $(b_k)$ (otherwise the left-side subsidiary node would be an open circle). However in anticipation of later considerations, it is interesting to look at this case. Then the corresponding $\mathcal{T}=\ket{\theta_\perp^k}\bra{\theta_\perp^k}$ and thus the resulting $\mathcal{T}^\prime$ is just the identity. This is correct: we have obtained a set of clause projectors for the subsidiary problem that includes the identity, and $\mathcal{I}\ket{s}=0$ can never be satisfied, indicating that we have reached a quantum problem with no solution (reassuring since the open circle denotes a classical \small{FALSE} outcome for the overall proposition).

Finally consider those clauses in the upper-tier \small{3-SAT} which do involve boolean $b_k$, and which `want' this boolean to be \small{FALSE}, i.e. they involve ``$\lnot b_k$''. We denote this subset of quantum projectors for the patent problem as $\mathcal{F}$. This is simply the converse of the case we have just considered: The clauses are not inherited by the right-side root, while the left-side root (where we have set $b_k\rightarrow$\small{TRUE}) does inherit the clauses with the simplification that $\lnot b_k$ is removed. The corresponding quantum operators for the lower-tier \small{3-SAT} are:
\[
\{\mathcal{F}^\prime\}\ \ \ \text{such that}\ \ \ \ \mathcal{F}_i^\prime\equiv\bra{{\bar\theta}^k_\perp}\ \mathcal{F}_i\ \ket{{\bar\theta}^k_\perp}.
\]
Analogous remarks to those above regarding the consequences of a clause $(b_k)$ associated with the parent, now apply here  when we consider $(\lnot b_k)$.

In summary: The set of quantum operators for the upper node is formed of three non-overlapping subsets,
\[
\{\mathcal{P}^U\}=\{\mathcal{X}\}\cup\{\mathcal{T}\}\cup\{\mathcal{F}\}.
\]
Meanwhile for the lower-left \small{3-SAT}, the corresponding quantum problem has projector set
\begin{equation}
\{\mathcal{P}^{L}\}=\{\mathcal{X}\}\cup\{\mathcal{F}^\prime\},
\label{eqn:leftP}
\end{equation}
and for the lower-right \small{3-SAT}, the corresponding quantum problem has projector set
\begin{equation}
\{\mathcal{P}^{R}\}=\{\mathcal{X}\}\cup\{\mathcal{T}^\prime\}.
\label{eqn:rightP}
\end{equation}

We will now consider the number of quantum solutions for the upper node. This quantum problem involves $n+1-k$ qubits, with indices $k$ to $n$ inclusive. In general any state of these qubits can be written as 
\begin{equation}
\ket{s}=\alpha\ket{\theta^k}\ket{\psi_1}+\beta\ket{\bar{\theta}^k}\ket{\psi_2}.\label{generalSoln}
\end{equation}
Here $\ket{\psi_1}$ and $\ket{\psi_2}$ are states of the qubits which have indices from $k+1$ to $n$ inclusive. In general this is a non-orthogonal decomposition since $\ket{\theta^k}$ and $\ket{\bar\theta^k}$ are independent but not orthogonal (we are considering the range $0<\theta<\tfrac{\pi}{2}$). Thus we may find $|\alpha|^2+|\beta|^2\neq 1$ but this is irrelevant to our analysis.

Now for $\ket{s}$ to be a solution to the quantum problem corresponding to the upper-tier node, we demand 
\begin{equation}
\mathcal{P}^U_i\ket{s}=0\ \ \ \forall i
\label{eqn:demandForUpper}
\end{equation}
which implies
\begin{eqnarray}
\mathcal{X}\ket{s}&=&0\ \ \ \forall \mathcal{X}\in\{\mathcal{X}\},\nonumber\\
\text{and}\ \ \ \mathcal{T}\ket{s}&=&0\ \ \ \forall \mathcal{T}\in\{\mathcal{T}\},\nonumber\\
\text{and}\ \ \ \mathcal{F}\ket{s}&=&0\ \ \ \forall \mathcal{F}\in\{\mathcal{F}\}.\nonumber
\end{eqnarray}
But substituting in our expression for $\ket{s}$,
\[
\mathcal{X}\ket{s}=\alpha\ket{\theta^k}\mathcal{X}\ket{\psi_1}+\beta\ket{\bar{\theta}^k}\mathcal{X}\ket{\psi_2}
\]
since $\mathcal{X}$ does not operate on the $k^\text{th}$ qubit. Since $\ket{\theta^k}$ and $\ket{\bar\theta^k}$ are distinct, the only way to satisfy $\mathcal{X}\ket{s}=0$ is for us to separately demand that
\[
\alpha\mathcal{X}\ket{\psi_1}=0\ \ \ \text{and}\ \ \ \ \beta\mathcal{X}\ket{\psi_2}=0.
\]
Meanwhile, substituting $\ket{s}$ on the second line yields
\[
\mathcal{T}\ket{s}= 0+\beta\mathcal{T}\ket{\bar{\theta}^k}\ket{\psi_2}.  
\]
because $\mathcal{T}$ involves $\ket{\theta_\perp^k}\bra{\theta_\perp^k}$. In fact,
\[
\mathcal{T}\ket{s}=\beta\ket{\theta_\perp^k}\braket{\theta_\perp^k}{\bar{\theta}^k}\,\mathcal{T}^\prime\ket{\psi_2}.  
\]
However, $\braket{\theta_\perp^k}{\bar{\theta}^k}$ is non-zero, so the condition that $\mathcal{T}\ket{s}=0$ means that 
\[
\beta\mathcal{T}^\prime\ket{\psi_2}=0.
\]
Similarly, substituting $\ket{s}$ into our third line yields the conclusion that 
\[
\alpha\mathcal{F}^\prime\ket{\psi_1}=0.
\]

If $\alpha\neq 0$, i.e. if the $\ket{\psi_1}$ term exists in $\ket{s}$, then we can summarise the constraints on $\ket{\psi_1}$ thus:
\begin{equation}
\mathcal{X}\ket{\psi_1}=0\ \  \forall \mathcal{X}\in\{\mathcal{X}\},\ \ \text{and}\ \ \ \ \mathcal{F}^\prime\ket{\psi_1}=0\ \ \forall \mathcal{F^\prime}\in\{\mathcal{F^\prime}\}. \label{leftSubset}
\end{equation}
We notice by glancing back to Eqn.~(\ref{eqn:leftP}) that these are exactly the conditions for a quantum solution to exist, for the problem corresponding to the left-root subsidiary problem. Similarly, if $\beta\neq 0$ then the conditions on $\ket{\psi_2}$ are
\begin{equation}
\mathcal{X}\ket{\psi_2}=0\ \ \forall \mathcal{X}\in\{\mathcal{X}\},\ \ \ \text{and}\ \ \ \ \ \mathcal{T}^\prime\ket{\psi_2}=0\ \ \forall \mathcal{T^\prime}\in\{\mathcal{T^\prime}\}. \label{rightSubset}
\end{equation}
Referring back to to Eqn.~(\ref{eqn:rightP}) we confirm these are exactly the conditions for a quantum solution in the problem corresponding to the right-root subsidiary problem.

Consequently, if there is no solution to the quantum problem corresponding to the left-root, nor the quantum problem corresponding to the right-root, then no satisfactory states $\ket{\psi_1}$ or $\ket{\psi_2}$ exist. We conclude that we cannot in fact satisfy Eqn.~(\ref{eqn:demandForUpper}) and thus there is no quantum solution to the upper tier problem.

However if there is no solution to the quantum problem corresponding to the left-root (i.e. no valid $\ket{\psi_1}$), but there are a finite number $N$ of quantum solutions associated with the other root (i.e. $N$ independent states can be selected for $\ket{\psi_2}$), then the parent node will have the $N$ quantum solutions of the form $\ket{s}=\ket{\bar\theta^k}\ket{\psi_2}$. The analogous principle applies to the mirror situation where only the left-root has a finite number of solutions.

In general, if the two roots are associated with numbers of quantum solutions $M$ and $N$ respectively, then the parent node will have $M+N$ solutions. This is the rule depicted in the central inset panel in Fig.\,\ref{fig:3sat_quantumCase}.

It remains only to consider the special cases that occur at the terminating bases of the root structure, where the entire \small{3-SAT} directly resolves to \small{TRUE} or \small{FALSE} (in the figures, these cases are denoted by the ticked square, and the  open circle, respectively).

Consider first the case where both roots of a node evaluate directly to \small{FALSE}. This is case (i) in Fig.\,\ref{fig:3sat_quantumCase}.
In order for the parent \small{3-SAT} to have the property that setting boolean $b_k$ to either \small{TRUE} or \small{FALSE} will, in both cases, result in an isolated clause evaluating to \small{FALSE}, it is necessary that the clause $(b_k)$ and the clause $(\lnot b_k)$ are both present. Note our `in isolation' condition: If, say, clause $(b_k)$ is not present but instead both $(b_k\lor b_j)$ and $(b_k\lor \lnot b_j)$ are present for some $j\neq k$, then we can of course infer that the setting $b_k\rightarrow$\small{FALSE} must ultimately lead to one of these clauses being \small{FALSE} -- however neither clause yet evaluates to \small{FALSE} when considered in isolation, so following the rules for constructing our tree, this right-side daughter will be a filled circle. Thus we see that indeed the single-boolean clauses $(b_k)$ and $(\lnot b_k)$ must both exist in the parents set of clauses. In the corresponding quantum problem, the projector corresponding to classical clause $(b_k)$ is ${\mathcal T}=\ket{\theta^k_\perp}\bra{\theta^k_\perp}$. Meanwhile the projector corresponding to classical clause $(\lnot b_k)$ is ${\mathcal F}=\ket{\bar \theta^k_\perp}\bra{\bar\theta^k_\perp}$. Recall that for a quantum solution $\ket{s}$ to exist, we require
\[
{\mathcal T}\ket{s}={\mathcal F}\ket{s}=0\ \ \ \text{(as with all other projectors)}.
\]
However the two projectors which we have just identified are non-identical (recall that we have $\theta>0$) and they each operate solely on the $k^\text{th}$ qubit. Thus there is no state $\ket{s}$ of the register of qubits that can simultaneously satisfy ${\mathcal T}\ket{s}={\mathcal F}\ket{s}=0$. The requirement is equivalent to asking a single qubit to be orthogonal to two distinct states simultaneously. We conclude that the patent node has no quantum solutions.

In the second special case (ii), we suppose that the right-side subsidiary \small{3-SAT} evaluates to \small{FALSE}. However the left-side root is such that its quantum analogue exists and has $N$ solutions (where $N=0$ is possible). Because the quantum problem exists, we infer that there is at least one remaining unspecified boolean $b_i$ in the subsidiary problems (in order that there is at least one qubit). We can ask, how many quantum solutions are associated with the open-circle node, i.e. how many independent states $\ket{\psi_2}$ meet the condition Eqn\,(\ref{leftSubset})? But we know that $(b_k)$ must be a clause in the parent \small{3-SAT}, and following the earlier reasoning this implies an identity operator in the set $\{\mathcal{T}^\prime\}$ so we conclude that the subsidiary quantum problem no solutions. Then, the parent node has $N+0=N$ quantum solutions.

For case (iii), first consider the simple case that $b_k$ is the final boolean to be assigned a value, i.e. $k=n$, the number of bits in the apex \small{3-SAT}. Then the pattern implies that the parent \small{3-SAT} has only one clause: $(b_n)$. The corresponding single quantum projector defines the state of the sole qubit, i.e. there is exactly one quantum solution. Now suppose that we are at a higher tier, i.e. $b_k$ is not the last boolean to be assigned a value. The number of quantum solutions contributed by the open-circle subsidiary node is still zero, for the same reason as in case (ii). How many quantum solutions are associated with the ticked node? Note that every clause in the parent \small{3-SAT} must include $b_k$, i.e. every clause must `want' boolean $b_k$ to have the value \small{TRUE}. Only in this case will the assignment $b_k\rightarrow$\,\small{TRUE} result in all clauses, individually, becoming \small{TRUE}. Therefore in the quantum problem associated with the parent node, the sets $\{\mathcal{X}\}$ and $\{\mathcal{F}\}$ are empty; all clause projectors belong to set $\{\mathcal{T}\}$. Then the term $\ket{\psi_1}$ in Eqn.\,(\ref{leftSubset}) is under no constraint at all since $\{\mathcal{X}\}$ and $\{\mathcal{F^\prime}\}$ are empty. It has a number of independent solutions equal to $2^q$ where $q$ is the number of qubits in $\ket{\psi_1}$. Notice that entering $q=0$ will correctly assert that there is one quantum solution for the parent node when there are no qubits at all in the subsidiary nodes (the case we considered initially).

It may seem necessary to discuss the case labelled as (iv) in the Figure, however in fact this patten will never occur. It would require every clause to be of the form $b_k\lor \lnot b_k$ or $b_k\lor \lnot b_k\lor b_i$ and such clauses are illegal in our prescription for \small{3-SAT}s since they contain a boolean in both direct and negated form, and are thus trivially true. Notice that if the apex \small{3-SAT} has no such clauses then neither can any of the lower tier \small{3-SAT}s (clauses do not gain variables as we descend the root).

Figure~\ref{fig:3sat_quantumCase} shows how these rules allow one to determine the number of quantum solutions associated with any node. Realising that the number of classical solutions follows the same rule, thus we have our result: If the number of solutions to the classical \small{3-SAT} is $n_S$, where $n_S$ may be zero, then the quantum system will eventually tend (as we repeatedly post-select on passing clause checks) to a state in an $n_S$ dimensional subspace. This subspace is spanned by the states
\[
\ket{\bm \theta}^j=\bigotimes_i \ R_Y\left(L_i^j\theta\right)\ket{+}_i \ \ \ j=1..n_S
\]
with $L_i^j=+1$ or $-1$ according to the truth value of boolean $i$ in the $j^{th}$ satisfying solution to the classical \small{3-SAT}.

\section{Derivation of lower bound on Hamiltonian gap}
\label{appendix:gapDerivation}

We begin by considering a Hamiltonian $\mathcal{H} = \frac{1}{m} \sum_i \mathcal{P}_i$. In Ref.~\cite{Liming2017} it was shown that by measuring $k$ random terms from $\mathcal{H}$, if each measurement results in the lowest energy state for that term, the resulting state $\rho^{(k)}$ of the system will be an approximation to the ground state of $\mathcal{H}$ satisfying
\[
\text{Tr}\left(P_\text{gs} \rho^{(k)} \right) \geq \left(1+(1-g(\mathcal{H}))^k \left(\frac{1}{\text{Tr}\left(P_\text{gs} \rho\right)} - 1\right)\right)^{-1}.
\]
In the above equation $g(\mathcal{H})$ is the energy gap between the ground state and first excited state of $\mathcal{H}$, $P_\text{gs}$ is the projector onto the ground state of $\mathcal{H}$ and $\rho$ is the initial state of the system. This can be approximated as 
\[
\text{Tr}\left(P_\text{gs} \rho^{(k)} \right) \geq 1 - e^{-k g(\mathcal{H})} \left(\frac{1}{\text{Tr}\left(P_\text{gs} \rho\right)} - 1\right).
\]
Hence a good approximation to the ground state is obtained whenever $e^{-k g(\mathcal{H})} \ll \text{Tr}\left(P_\text{gs} \rho\right)$. In the context of the quantum 3-SAT solver described in the main text, $\text{Tr}\left(P_\text{gs} \rho\right)$ can be easily computed as $\cos\left(\frac{\theta}{2}\right)^n$. In order to bound $k$ it is then only necessary to bound $g(\mathcal{H})$.

Since $\mathcal{H}$ has a ground state energy of zero by construction, $g(\mathcal{H})$ is given by the energy of the first excited state which can be obtained by minimising $\bra{\Phi}\mathcal{H}\ket{\Phi}$ over states $\ket{\Phi}$ which are orthogonal to the null space of $\mathcal{H}$. We can $\ket{\Phi}$ using a non-orthogonal basis given by
\[
\{\ket{\theta_x} = \bigotimes_{i=1}^n R_Y\left((-1)^{x_i}\theta\right)\ket{+} ~| x \in \{0,1\}^n\}.
\]
This yields $\ket{\Phi} = \sum_x \alpha_x \ket{\theta_x}$. A bound on $\bra{\Phi}\mathcal{H}\ket{\Phi}$ can then be obtained as follows.

\begin{align*}
\bra{\Phi}\mathcal{H}\ket{\Phi} &= \frac{1}{m} \sum_i \bra{\Phi} \mathcal{P}_i \ket{\Phi}\\
&= \frac{1}{m} \sum_i \sum_{x,y} \alpha_x^* \alpha_y \bra{\theta_x} \mathcal{P}_i \ket{\theta_y}.
\end{align*}
Note that $\bra{\theta_x} \mathcal{P}_i \ket{\theta_y}$ is either equal to zero or to $sin(\theta)^6$, with the latter occurring when the three bits of $x$ and $y$ on which the projector in $\mathcal{P}_i$ acts non-trivially are equal and are equal to the exact 3-bit string that violates the $i$th clause of the 3-SAT instance. For this reason we will sum $x$ and $y$ only over the strings $S_i$ which violate the $i$the clause. This yields
\begin{align*}
\bra{\Phi}\mathcal{H}\ket{\Phi} &= \frac{1}{m} \sum_i \sum_{x,y\in S_i} \alpha_x^* \alpha_y \bra{\theta_x} \mathcal{P}_i \ket{\theta_y} \\
&= \frac{1}{m} \sum_i \sum_{x,y\in S_i} \alpha_x^* \alpha_y \sin(\theta)^6 \cos(\theta)^{D_{xy}},
\end{align*}
where $D_{xy}$ is the Hamming distance between $x$ and $y$. It is convenient to re-express this summation as 
\begin{align*}
\bra{\Phi}\mathcal{H}\ket{\Phi} &= \frac{1}{m} \sum_i \sum_{x,y\in S_i} \alpha_x^* \alpha_y \sin(\theta)^6 \bra{x} M^{\otimes n} \ket{y},
\end{align*}
where $\ket{x}$ and $\ket{y}$ denote the vectors in the standard basis and $M=\begin{bmatrix}
    1 & \cos(\theta) \\
    \cos(\theta) & 1
\end{bmatrix}$.

The expectation value of $\mathcal{H}$ can then be expressed as 
\begin{align*}
\bra{\Phi}\mathcal{H}\ket{\Phi} &= \sin(\theta)^6 \text{Tr}\left(\left(\frac{1}{m} \sum_i \sum_{x,y\in S_i} \alpha_x^* \alpha_y \ket{y}\bra{x}\right) M^{\otimes n}\right),
\end{align*}
which can be bounded from below using the matrix inequality $\text{Tr}\left(AB\right) \geq \text{Tr}(A) \lambda_\text{min}(B)$, where $\lambda_\text{min}(B)$ is the minimum eigenvalue of $B$, to give 
\begin{align*}
\bra{\Phi}\mathcal{H}\ket{\Phi} &\geq \sin(\theta)^6 \lambda_\text{min}(M)^n \text{Tr}\left(\frac{1}{m} \sum_i \sum_{x,y\in S_i} \alpha_x^* \alpha_y \ket{y}\bra{x}\right) \\
&= \frac{1}{m}\sin(\theta)^6 \lambda_\text{min}(M)^n \sum_i \sum_{x\in S_i} |\alpha_x|^2.
\end{align*}
To evaluate the above equation, we first note that $\lambda_\text{min}(M)=1 - \cos(\theta)$. We then note that each $x$ which is not a solution to the 3-SAT instance must appear in at least one of the sets $S_i$. Hence we have the inequality 
\begin{align*}
\bra{\Phi}\mathcal{H}\ket{\Phi} &\geq \frac{1}{m}  \sin(\theta)^6 (1-\cos(\theta))^n \sum_{x\notin S} |\alpha_x|^2,
\end{align*}
where $S$ is the set of solutions to the 3-SAT instance.

It remains to bound $\sum_{x\notin S} |\alpha_x|^2$. To do this we note that $1 = \braket{\Phi}{\Phi} = \sum_{x,y \notin S} \alpha_x^* \alpha_y \braket{\theta_x}{\theta_y} - \sum_{u,v \in S} \alpha_u^* \alpha_v \braket{\theta_u}{\theta_v}$ since $\braket{\Phi}{\theta_s} = 0$ for all $s \in S$. Hence
\begin{align*}
	1 &\leq \sum_{x,y \notin S} \alpha_x^* \alpha_y \braket{\theta_x}{\theta_y}\\
	&= \sum_{x,y \notin S} \alpha_x^* \alpha_y \bra{x}M^n\ket{y}\\
	&\leq \lambda_\text{max}(M)^n \left|\left|\sum_x \alpha_x \ket{x}\right|\right|^2_2,
\end{align*}
where $\lambda_\text{max}(M)$ is the maximum eigenvalue of $M$. Hence $\sum_{x\notin S} |\alpha_x|^2 \geq (1+\cos(\theta))^{-n}$ and we obtain
\begin{align*}
\bra{\Phi}\mathcal{H}\ket{\Phi} &\geq \frac{1}{m}  \sin(\theta)^6 \left(\frac{1-\cos(\theta)}{1+\cos(\theta)}\right)^n.
\end{align*}

\end{document}